\documentclass[format=acmsmall, review=false, screen=true]{acmart}

\usepackage{graphicx}
\usepackage{balance}
\usepackage{xcolor}
\usepackage{url}
\usepackage{amssymb}
\usepackage{amsmath}
\usepackage{multirow}
\usepackage{booktabs}
\usepackage{makecell}
\usepackage{natbib}
\usepackage[caption=false]{subfig}
\usepackage{wrapfig}
\usepackage{xcolor, soul}
\usepackage{siunitx}
\usepackage{enumitem}
\usepackage{hyperref}
\usepackage{rotating}
\usepackage{tablefootnote}
\usepackage[noend]{algpseudocode}
\usepackage[linesnumbered
			,vlined]{algorithm2e}

\definecolor{DarkGray}{gray}{.45}

\newcommand{\parag}[1]{\vspace{0.2cm}\noindent\textbf{\textsf{#1.}}}

\newcommand{\mytablescale}{0.9}

\SetKwIF{If}{ElseIf}{Else}{{\textsf{if}}}{}{\textsf{else if}}{{\textsf{else :}}}{}
\SetKwFor{For}{\textsf{for}}{}{}
\SetKwFor{While}{\textsf{while}}{}{}
\newcommand{\code}[1]{\textsf{\textbf{#1}}}
\newcommand{\func}[1]{\textsf{{#1}}}

\newcommand{\bit}[1]{\texttt{#1}}
\newcommand{\enc}[3]{\textsf{#1}_{#3}(#2)}
\newcommand{\B}{\texttt{bin}}

\usepackage{xcolor, soul}

\newcommand{\gray}[1]{{\color{gray}{#1}}}

\usepackage{pbox}
\usepackage{makecell}
\usepackage{tikz}
\usetikzlibrary{calc}

\newcommand{\total}{\textsf{GiB}}
\newcommand{\bpi}{\textsf{bits/int}}
\newcommand{\nspi}{\textsf{ns/int}}
\newcommand{\docid}{docID}

\newcommand{\gov}{\textsf{Gov2}}
\newcommand{\clue}{\textsf{ClueWeb09}}
\newcommand{\cc}{\textsf{CCNews}}

\newcommand{\RNum}[1]{\uppercase\expandafter{\romannumeral #1\relax}}

\newcommand{\seq}{\mathcal{S}}
\newcommand{\pr}{\mathbb{P}}

\newcommand{\AND}{\textup{\textsf{AND}}}
\newcommand{\OR}{\textup{\textsf{OR}}}

\newcommand{\select}{\textup{\textsf{Select}}}

\newcommand{\rank}{\textup{\textsf{Rank}}}
\newcommand{\access}{\textup{\textsf{Access}}}
\newcommand{\Nextgeq}{\textup{\textsf{NextGEQ}}}

\newcommand{\suc}{\textup{\textsf{Successor}}}

\newcommand{\method}[1]{{\sf{#1}}}

\newcommand{\var}[1]{\mbox{\emph{#1}}}

\newcommand{\interp}{{\method{BIC}}}
\newcommand{\vb}{{\method{VByte}}}
\newcommand{\opt}{{\method{PEF}}}
\newcommand{\roaropt}{{\method{Roaring}}}
\newcommand{\slice}{{\method{Slicing}}}
\newcommand{\simples}{{\method{Simple16}}}

\newcommand{\bpt}{{\method{BP32}}}
\newcommand{\bpsimd}{{\method{SIMD-BP128}}}
\newcommand{\cdelta}{{$\delta$}}

\newcommand{\rice}{{\method{Rice}}}

\newcommand{\bic}{{\method{BIC}}}
\newcommand{\ef}{{\method{EF}}}
\newcommand{\pef}{{\method{PEF}}}
\newcommand{\qmx}{\method{QMX}}
\newcommand{\dint}{\method{DINT}}

\newcommand{\ans}{{\method{ANS}}}
\newcommand{\varigiu}{{\method{Varint-G8IU}}}
\newcommand{\varigb}{{\method{Varint-GB}}}
\newcommand{\vbyte}{{\method{VByte}}}
\newcommand{\optvb}{{\method{Opt-VByte}}}
\newcommand{\masked}{{\method{Masked-VByte}}}
\newcommand{\stream}{{\method{Stream-VByte}}}
\newcommand{\simplen}{{\method{Simple9}}}
\newcommand{\simplee}{{\method{Simple8b}}}

\newcommand{\pfor}{\method{PFor}}
\newcommand{\opf}{\method{Opt-PFor}}

\acmJournal{CSUR}
\setcopyright{acmlicensed}
\begin{document}

\title[Techniques for Inverted Index Compression]{Techniques for Inverted Index Compression}

\author{Giulio Ermanno Pibiri}
\affiliation{%
 \institution{ISTI-CNR}
 \streetaddress{Via Giuseppe Moruzzi 1, 56124}
 \city{Pisa}
 \country{Italy}}

\author{Rossano Venturini}
\affiliation{%
  \institution{University of Pisa}
  \streetaddress{Largo Bruno Pontecorvo 3, 56127}
  \city{Pisa}
  \country{Italy}}

\email{giulio.ermanno.pibiri@isti.cnr.it}
\email{rossano.venturini@unipi.it}

\begin{abstract}
The data structure at the core of large-scale search engines is the \emph{inverted index},
which is essentially a collection of sorted integer sequences called \emph{inverted lists}.
Because of the many documents indexed by such engines and stringent performance requirements
imposed by the heavy load of queries, the inverted index stores billions of integers
that must be searched efficiently.
In this scenario, \emph{index compression} is essential because
it leads to a better exploitation of the computer memory hierarchy for faster
query processing and, at the same time, allows reducing
the number of
storage machines.

The aim of this article is twofold: first, surveying the
encoding algorithms suitable for inverted index compression and, second,
characterizing the performance of the inverted index through experimentation.
\end{abstract}

\ccsdesc[500]{Information systems~Search index compression}
\ccsdesc[500]{Information systems~Search engine indexing}

\keywords{Inverted Indexes; Data Compression; Efficiency}

\maketitle

\renewcommand{\shortauthors}{G.\,E. Pibiri and R. Venturini}

\section{Introduction}\label{sec:intro}

Consider a collection of textual documents each described, for this purpose,
as a set of terms.
For each distinct term $t$ appearing in the collection,
an integer sequence $\seq_t$ is built and lists, in sorted order, all the identifiers of the documents
(henceforth, {\docid}s) where the term appears.
The sequence $\seq_t$ is called the \emph{inverted list}, or \emph{posting list},
of the term $t$
and the set of inverted lists for all the distinct terms
is the subject of this article
-- the data structure known as the \emph{inverted index}.
Inverted indexes can store additional information about each term,
such as the set of positions where the terms appear in the documents (in \emph{positional} indexes)
and the number of occurrences of the terms in the documents, i.e., their \emph{frequencies}~\cite{ZobelM06,Raghavan-book,Buttcher-book}.
In this article we consider the {\docid}-\emph{sorted} version of the inverted index and we ignore additional
information about each term.
The inverted index is the data structure at the core of large-scale search engines, social networks and storage architectures~\cite{ZobelM06,Raghavan-book,Buttcher-book}.
In a typical use case, it is used to index millions of documents, resulting
in several billions of integers.
We mention some noticeable examples.

Classically, inverted indexes are used to support full-text search in databases~\cite{Raghavan-book}.
Identifying a set of documents containing \emph{all} the terms in a user query reduces to the problem of \emph{intersecting} the inverted lists associated to the terms in the query.
Likewise, an inverted list can be associated to a user in a social network (e.g., Facebook)
and stores the sequence of all friend identifiers of the user~\cite{2013:curtiss.becker.ea}.
Database systems based on SQL often precompute the list of row identifiers matching a specific frequent predicate over a large table, in order to speed up the execution of a query involving the conjunction of many predicates~\cite{hristidis2003efficient,raman2007lazy}.
Key-value storage is a popular database design principle, adopted by architectures such as
Apache Ignite, Redis, InfinityDB, BerkeleyDB and many others.
Common to all such architectures is the organization of data elements falling into the same bucket
due to an hash collision: the list of all such elements is materialized,
which is essentially an inverted list~\cite{debnath2011skimpystash}.

Because of the huge quantity of indexed documents and heavy query loads,
\emph{compressing} the inverted index is indispensable because
it can introduce a twofold advantage
over a non-compressed representation:
feed faster memory levels with more data and,
\emph{hence}, speed up the query processing algorithms.
As a result, the design of techniques that compress the index effectively
while maintaining a noticeable decoding speed is a well-studied
problem,
that dates back to more than 50 years ago, and still a very active field of research.
In fact, many representation for inverted lists are known,
each exposing a different space/time trade-off:
refer to the timeline shown in Table~\ref{tab:timeline} and references therein.


\begin{table}
\captionsetup[subfloat]{labelformat=empty}
\caption{Timeline of techniques.
\label{tab:timeline}}
\vspace{-0.5cm}
\subfloat[]{
    \scalebox{\mytablescale}{
    	\begin{tabular}{l l}
\toprule

%

1949 & Shannon-Fano~\cite{Shannon48,Fano49} \\
\cmidrule(lr){1-2}
1952 & Huffman~\cite{Huffman52} \\
\cmidrule(lr){1-2}
1963 & Arithmetic~\cite{abramson1963information}\tablefootnote{Actually somewhat before 1963. See Note 1 on page 61 in the book by~\citet*{abramson1963information}.} \\
\cmidrule(lr){1-2}
1966 & Golomb~\cite{Golomb66} \\
\cmidrule(lr){1-2}
1971 & Elias-Fano~\cite{Fano71,Elias74}; Rice~\cite{Rice71} \\
\cmidrule(lr){1-2}
1972 & Variable-Byte and Nibble~\cite{thiel1972program} \\
\cmidrule(lr){1-2}
1975 & Gamma and Delta~\cite{elias75} \\
\cmidrule(lr){1-2}
1978 & Exponential Golomb~\cite{teuhola1978compression} \\
\cmidrule(lr){1-2}
1985 & Fibonacci-based~\cite{fraenkel1985robust,apostolico1987robust} \\
\cmidrule(lr){1-2}
1986 & Hierarchical bit-vectors~\cite{fraenkel1986improved} \\
\cmidrule(lr){1-2}
1988 & Based on Front Coding~\cite{cfk88sigir} \\
\cmidrule(lr){1-2}
1996 & Interpolative~\cite{1996:moffat.stuiver,2000:moffat.stuiver} \\
\cmidrule(lr){1-2}
\multirow{2}{*}{1998} & Frame-of-Reference (For)~\cite{GoldsteinRS98}; \\
& modified Rice~\cite{Vo:1998:CIF:290941.291011} \\
\cmidrule(lr){1-2}
2003 & SC-dense~\cite{brisaboa2003s} \\
\cmidrule(lr){1-2}
2004 & Zeta~\cite{boldi2004webgraph,2005:boldi-vigna} \\

\bottomrule
\end{tabular}

    }
}
\subfloat[]{
    \scalebox{\mytablescale}{
    	\begin{tabular}{l l}
\toprule

%

\multirow{2}{*}{2005} & Simple-9, Relative-10, and Carryover-12~\cite{AnhM05}; \\
& RBUC~\cite{moffat2005binary} \\
\cmidrule(lr){1-2}
2006 & PForDelta~\cite{zukowski06super}; BASC~\cite{moffat2006binary} \\
\cmidrule(lr){1-2}
2008 & Simple-16~\cite{zlt08www}; Tournament~\cite{teuhola2008tournament} \\
\cmidrule(lr){1-2}
2009 & ANS~\cite{duda09}; Varint-GB~\cite{2009:dean}; Opt-PFor~\cite{2009:yan.ding.ea} \\
\cmidrule(lr){1-2}
2010 & Simple8b~\cite{AnhM10}; VSE~\cite{2010:silvestri.venturini}; SIMD-Gamma~\cite{schlegel2010fast} \\
\cmidrule(lr){1-2}
2011 & Varint-G8IU~\cite{2011:stepanov.gangolli.ea}; Parallel-PFor~\cite{ao2011efficient} \\
\cmidrule(lr){1-2}
2013 & DAC~\cite{brisaboa2013dacs}; Quasi-Succinct~\cite{2013:vigna} \\
\cmidrule(lr){1-2}
\multirow{2}{*}{2014} & partitioned Elias-Fano~\cite{2014:ottaviano.venturini}; QMX~\cite{2014:trotman}; \\
& Roaring~\cite{chambi2016better,lemire2016consistently,lemire2018roaring} \\
\cmidrule(lr){1-2}
\multirow{2}{*}{2015} & BP32, SIMD-BP128, and SIMD-FastPFor~\cite{2013:lemire.boytsov}; \\
& Masked-VByte~\cite{2015:plaisance.kurz.ea} \\
\cmidrule(lr){1-2}
2017 & clustered Elias-Fano~\cite{2017:pibiri.venturini} \\
\cmidrule(lr){1-2}
\multirow{3}{*}{2018} & Stream-VByte~\cite{2018:lemire.kurz.ea}; ANS-based~\cite{ANS1,ANS2}; \\
& Opt-VByte~\cite{TKDE19}; SIMD-Delta~\cite{trotman2018elias}; \\
& general-purpose compression libraries~\cite{petri2018compact}; \\
\cmidrule(lr){1-2}
2019 & DINT~\cite{DINT}; Slicing~\cite{slicing} \\

\bottomrule
\end{tabular}

    }
}
\end{table}

\parag{Organization}
We classify the techniques in a hierarchical manner by identifying
\emph{three} main classes.
The first class consists of algorithms that compress a single integer
(Section~\ref{sec:integer}).
The second class covers algorithms that compress many integers together,
namely an inverted list (Section~\ref{sec:list}).
The third class describes a family of algorithms that represent many lists together,
i.e., the whole inverted index (Section~\ref{sec:index}).
In our intention, this first part of the survey is devoted to readers
who are new to the field of integer compression.

This hierarchical division is natural and intuitive.
First, it reflects the flexibility of the algorithms, given that
algorithms in a higher class can be used
to represent the unit of compression of the algorithms in a lower class,
but not the other way round.
For example, an algorithm that compresses
a single integer at a time (first class) can obviously be used to represent
a list of integers (unit of compression of the second class) by just
concatenating the encoding of each single integer in the list.
Second, it shows that less flexibility can be exploited to enlarge
the ``visibility'' of the algorithms.
For example, algorithms in the second class seek opportunities
for better compression by looking for regions of similar integers
in the list. Instead, algorithms in the third class seek for such
regularities across many lists (or even across the whole index).

After the description of the techniques, we provide pointers
to further readings (Section~\ref{sec:further}).
The last part of the article is dedicated to 
the experimental comparison of the paradigms used to represent
the inverted lists (Section~\ref{sec:experiments}).
In our intention,
this part is targeted to more experienced
readers who are already familiar with the research field
and the practical implementation of the techniques.
We release the full experimental suite
at \url{https://github.com/jermp/2i_bench}, in the hope
of spurring further research in the field.
We conclude the survey by summarizing experimental lessons 
and discussing some future research directions (Section~\ref{sec:conclusions}).

\section{Integer Codes}\label{sec:integer}

The algorithms we consider in this section compress a single integer.
We first introduce some preliminary notions that help to better
illustrate such algorithms.
Let $x > 0$ indicate the integer to be represented\footnote{Throughout the section
we present the codes for positive integers. Also, we use 1-based indexes for arrays.
This is a deliberate choice for
illustrative purposes. The reader should be aware that some implementations
at \url{https://github.com/jermp/2i_bench} may differ from the content of the paper
and should be careful in comparing them.}.
Sometimes we assume
that an upper bound on the largest value that $x$ can take is known,
the so-called ``universe'' of representation,
and we indicate that with $U$. Therefore, it holds $x \leq U$, $\forall x$.
Let $\enc{C}{x}{}$ be the bit string encoding $x$ -- the \emph{codeword} of $x$ according
to the code \textsf{C} -- 
and $|\enc{C}{x}{}|$ its length in bits.

The most classical solution to the problem of integer coding
is to assign $x$ a \emph{uniquely-decodable} 
variable-length code, in order to decode without ambiguity (thus, correctly)
from left to right.
The aim of the algorithms we present in the following
is to assign the smallest codeword as possible.
In this regard, a distinction is made between the identification of a \emph{set of
codeword lengths} and the subsequent \emph{codeword assignment} phase,
that is the mapping from integer identifiers to binary strings.
Once the codeword lengths have been computed, the specific codewords are
actually irrelevant provided that no codeword is a prefix of another one
(prefix-free condition) in order to guarantee unique decodability --
a key observation widely documented in the literature~\cite{LH87,moffat1997implementation,huffman_coding,moffat2002compression,mg}.
Therefore, we can think of a code as being a (multi-) set of codeword lengths.
This interpretation has the advantage of being independent from the specific
codewords and makes possible to choose the assignment that is best
suited for fast decoding -- an important property as far as practicality is
concerned for inverted index compression.
Throughout this section we opt for a \emph{lexicographic assignment} of the codewords,
that is the codewords are in the same lexicographic order as the integers they
represent. This property will be exploited for fast decodability
in Section~\ref{sec:encoding_decoding}.
The crucial fact about these initial remarks is that \emph{all} prefix-free codes
can be arranged in this way, hence providing a
``canonical'' interface for their treatment.

Another key notion is represented by the \emph{Kraft}-\emph{McMillan inequality}~\cite{kraft1949device,mcmillan1956two}
that gives a necessary
and sufficient condition for the existence of a uniquely-decodable code for a given
set of codeword lengths, where no codeword is a prefix of another one.
More formally, it must hold
$\sum_{x=1}^u 2^{-|\enc{C}{x}{}|} \leq 1$
for the code to be uniquely-decodable.

It should be intuitive that no code is optimal for \emph{all}
possible integer distributions.
According to~\citet*{Shannon48}, the ideal codeword length
of the integer $x$ should be $\log_2(1/\pr(x))$ bits long, where $\pr(x)$ is
the probability of occurrence of $x$ in the input.
Therefore, by solving the equation
$|\enc{C}{x}{}| = \log_2(1/\pr(x))$ with respect to $\pr(x)$,
the distribution for which the considered integer code is optimal
can be derived.

Lastly, we also remark that sometimes it could be useful to implement a sub-optimal code if
it allows faster decoding for a given application, and/or to organize the output bit stream
in a way that is more suitable for special hardware instructions
such as SIMD (Single-Instruction-Multiple-Data)~\cite{SIMDIntel}.
SIMD is a computer organization that exploits the
independence of multiple data objects to execute a single instruction on these objects simultaneously.
Specifically, a single instruction is executed for every element
of a \emph{vector} -- a large(r) machine register that packs multiple elements together,
for a total of 128, 256, or even 512 bits of data.
SIMD is widely used to accelerate the execution of many data-intensive tasks,
and (usually) an optimizing compiler is able to automatically ``vectorize''
code snippets to make them run faster.
Many algorithms that we describe in this article exploit SIMD instructions.

\subsection{Encoding and decoding prefix-free codes}\label{sec:encoding_decoding}

We now illustrate how the lexicographic ordering of the codewords can be
exploited to achieve efficient encoding and decoding of prefix-free codes.
By ``efficiency'' we mean that a small and fixed number of instructions
is executed for each encoded/decoded integer,
thus avoiding the potentially expensive bit-by-bit processing.
The content of this section is based on the work by~\citet*{moffat1997implementation}
that also laid the foundations for (part of) Chapter 4 of their own book~\cite{moffat2002compression}
and the descriptions in Section 3.1 and 3.2 of the survey by~\citet*{huffman_coding}.

We are going to refer to Table~\ref{tab:encoding_decoding1} as an example code.
The table shows
the codewords assigned to the integers 1..8
by the \emph{gamma} code, the first non-trivial code we will describe later
in Section~\ref{subsec:gamma_delta}.
Let $M$ be the length of the longest codeword, that is $M = \max_x\{|\enc{C}{x}{}|\}$.
In our example, $M=7$.
The column headed \emph{lengths} reports the lengths (in bits) of the
codewords, whereas \emph{values} are the decimal representation
of the codewords seen as left-justified $M$-bit integers.
If such columns are represented
with two parallel arrays indexed by the $x$ value,
then the procedures for encoding/decoding
are easily derived as follows.
To encode $x$, just write to the output stream the \emph{most significant}
(from the left)
$\var{lengths}[x]$ bits of $\var{values}[x]$.
To decode $x$, we use a variable $\var{buffer}$ always holding
$M$ bits from the input stream.
The value assumed by \var{buffer} is, therefore, an $M$-bit integer
that we search in the array $\var{values}$ determining the codeword length
$\ell = \var{length}[x]$
such that $\var{values}[x] \leq \var{buffer} < \var{values}[x + 1]$.
Now that $\ell$ bits are consumed from $\var{buffer}$, other $\ell$ bits are
fetched from the input stream via a few masking and shifting operations.

\begin{table}
\caption{In (a), an example prefix-free code for the integers 1..8, along with
associated codewords, codeword lengths and corresponding left-justified, 7-bit, integers.
The codewords are left-justified to better highlight their lexicographic
order.
In (b), the compact version of the table in (a), used by the encoding/decoding
procedures coded in Fig.~\ref{alg:encoding_decoding}.
The ``values'' and ``first'' columns are padded with a sentinel value (in gray)
to let the search be well defined.
\label{tab:encoding_decoding}}
\subfloat[]{\scalebox{\mytablescale}{\begin{tabular}{clcc}
\toprule

$x$ & codewords & \emph{lengths} & \emph{values} \\

\midrule

1 & \bit{0} & 1 & 0 \\
2 & \bit{100} & 3 & 64 \\
3 & \bit{101} & 3 & 80 \\
4 & \bit{11000} & 5 &  96 \\
5 & \bit{11001} & 5 & 100 \\
6 & \bit{11010} & 5 & 104 \\
7 & \bit{11011} & 5 & 108 \\
8 & \bit{1110000} & 7 & 112 \\
-- & -- & -- & \gray{127} \\

\bottomrule
\end{tabular}
}
\label{tab:encoding_decoding1}}
\hspace{1cm}
\subfloat[]{\scalebox{\mytablescale}{\begin{tabular}{ccc}
\toprule

\emph{lengths} & \emph{first} & \emph{values} \\

\midrule

1 & 1  &  0  \\
{2} & {2}  &  {64}  \\
3 & 2  &  64 \\
{4} & {4}  &  {96}  \\
5 & 4  &  96 \\
{6} & {8}  &  {112} \\
7 & 8  & 112 \\
-- & \gray{9} & \gray{127} \\

\bottomrule
\end{tabular}
}
\label{tab:encoding_decoding2}}
\end{table}

For example, to encode the integer 4, we
omit the $\var{lengths}[4] = 5$ most significant bits from the binary representation
of $\var{values}[4] = 96$ as a 7-bit integer, that is \bit{1100000}.
The obtained codeword for 4 is, therefore, \bit{11000}.
Instead, assume we want to decode the next integer from a \var{buffer} configuration
of \bit{1010100}. This 7-bit integer is 84. By searching 84 in the \var{values} array,
we determine the index $x = 3$ as $80 = \var{values}[3] < 84 < \var{values}[4] = 96$.
Therefore, the decoded integer is 3 and we can fetch the next 
$\var{lengths}[3] = 3$ bits from the input stream.
(It is easy to see that the \var{buffer} configuration \bit{1010100}
holds the encoded values 3, 0, and 2.)

However, the cost of storing the \var{lengths} and \var{values} arrays
can be large because they can hold as many values as $U$.
The universe $U$
is typically in the range of tens of millions for a typical
inverted index benchmark (see also Table~\ref{tab:datasets} at page~\ref{tab:datasets}
for a concrete example), thus resulting in
large lookup tables that do not fit well in the computer cache hierarchy.
Fortunately enough, it is possible to replace such arrays with
two compact arrays of just $M+1$ integers each when the codewords
are assigned in lexicographic order.
Recall that we have defined $M$ to be the longest codeword length.
(In our example from Table~\ref{tab:encoding_decoding}, $M$ is 7.)
We expect to have
$M \leq U$ and, in particular, $M \ll U$, which is always valid in practice
unless $U$ is very small.


\begin{figure}[t]
\scalebox{\mytablescale}{\input{algs/encoding.tex}}
\scalebox{\mytablescale}{\input{algs/decoding.tex}}
\caption{Encoding and decoding procedures using two parallel arrays \var{first} and \var{values}
of $M+1$ values each.
\label{alg:encoding_decoding}}
\end{figure}

Table~\ref{tab:encoding_decoding2} shows the ``compact'' version of Table~\ref{tab:encoding_decoding1}, where other two arrays, \var{first} and \var{values},
of $M+1$ integers each are used. Both arrays are now indexed by codeword length $\ell$.
In particular, $\var{first}[\ell]$ is the first integer that is assigned a codeword of
length equal to $\ell$,
with $\var{values}[\ell]$ being the corresponding $M$-bit integer representation.
For example, $\var{first}[3] = 2$ because 2 is the first integer represented with a codeword
of length 3.
Note that not every possible codeword length could be used.
In our example, we are not using codewords of length 2, 4 and 6.
These unused lengths generate some ``holes'' in the \var{first} and \var{values} arrays.
A hole at position $\ell$ is then filled with the value corresponding to the
smallest codeword length $\ell^{\prime} > \ell$
that is used by the code.
For example, we have a hole at position $\ell = 2$ because we have no codeword of length 2.
Therefore, $\var{first}[2]$ and $\var{values}[2]$ are filled with 2 and 64 respectively
because these are the values corresponding to the codeword length $\ell^{\prime} = 3$.
With these two arrays it is possible to derive the compact pseudo code illustrated
in Fig.~\ref{alg:encoding_decoding}, whose correctness is left to the reader.
The function $\func{Write}(\var{val},\var{len})$ writes the \var{len} low bits of the
value \var{val} to the output stream;
conversely, the function $\func{Take}(\var{len})$ fetches the next \var{len}
bits from the input stream and interprets them as an integer value.
We now discuss some salient aspects of the pseudo code along with examples,
highlighting the benefit of working with lexicographically-sorted codewords.

Assigning lexicographic codewords
makes possible to use the offsets computed in line 3 of both listings in Fig.~\ref{alg:encoding_decoding} to perform encoding/decoding,
essentially allowing a sparse representation of the mapping from integers to codewords
and vice versa.
As an example, consider the encoding of the integer 6.
By searching the \var{first} array, we determine $\ell = 5$.
Now, the difference $\var{offset} = 6 - \var{first}[5] = 6 - 4 = 2$
indicates that 6 is the $(\var{offset}+1)$-th integer, the 3rd in this case,
that is assigned a codeword of length 5.
Therefore, starting from $\var{values}[5] = 96$
we can derive the $M$-bit integer corresponding to the encoding of 6
using such offset, via $96 + 2 \times 4 = 104$.
It is easy to see that this computation is correct only when
the codewords are assigned lexicographically, otherwise it would not be possible
to derive the $M$-bit integer holding the codeword.
The same reasoning applies when considering the decoding algorithm.

Another advantage of working with ordered codewords is that
both \var{first} and \var{values} arrays are sorted, thus binary search can be
employed to implement the identification of $\ell$ at line 2 of both
encoding and decoding algorithms.
This step is indeed the most expensive one.
Linear search could result faster than binary search for its cache-friendly nature,
especially when $M$ is small.
If we are especially concerned with decoding efficiency and can trade a bit more
working space, it is possible to identify $\ell$ via direct addressing, i.e.,
in $O(1)$ per decoded symbol,
using a $2^M$-element table indexed by \var{buffer}.
Other options are possible, including hybrid strategies using a blend of search
and direct addressing:
for a discussion of these options, again refer to the work
by~\citet*{moffat1997implementation,moffat2002compression},
~\citet*[Section 3.1 and 3.2]{huffman_coding}, and references therein.

\subsection{Unary and Binary}\label{sec:unary_binary}

Perhaps the most primitive form of integer representation is \emph{unary} coding,
that is the integer $x$ is represented by a run of $x-1$ ones plus a final zero:
$\bit{1}^{x-1}\bit{0}$.
The presence of the final \bit{0} bit implies that the encoding is uniquely-decodable:
decoding bit-by-bit will keep reading ones until we hit a \bit{0} and then report
the number of read ones by summing one to this quantity.
Because we need $x$ bits to represent the integer $x$, that is $|\enc{U}{x}{}| = x$,
this encoding strongly favours small integers.
For example, we can represent 2 with just 2 bits (\bit{10}) but we would need
503 bits to represent the integer 503.
Solving $\lceil \log_2(1 / \pr(x)) \rceil = x$ yields that the unary
code is optimal whenever $\pr(x) = 1/2^x$.

We indicate with $\B(x,k)$ the \emph{binary representation} of an integer $0 \leq x < 2^k$
using $k$ bits.
When we just write $\B(x)$,
it is assumed that $k = \lceil \log_2(x+1) \rceil$, which is the minimum number
of bits necessary to represent $x$.
We also distinguish between $\B(x,k)$ and the \emph{binary codeword} $\enc{B}{x}{}$
assigned to an integer $x > 0$.
Given that we consider positive integers only,
we use the convention that $\enc{B}{x}{}$ is $\B(x-1)$ throughout this section.
For example, $\enc{B}{6}{} = \B(5) = \bit{101}$.
See the second column of Table~\ref{tab:codes} for more examples.
This means that, for example, $|\enc{B}{4}{}|$ is 2 and not 3;
$\enc{B}{503}{}$ just needs $\lceil \log_2 503 \rceil = 9$ bits instead of 503 as needed
by its unary representation.
The problem of binary coding is that it is \emph{not} uniquely-decodable unless we know
the number of bits that we dedicate to the representation of \emph{each} integer in the coded stream.
For example, if the integers in the stream are drawn from a universe $U$ bounded by
$2^k$ for some $k > 0$, then each integer can be represented with
$\lceil \log_2 U \rceil \leq k$ bits, with an implied distribution of $\pr(x) = 1/2^k$.
(Many compressors that we present in Section~\ref{sec:list} exploit this simple strategy.)
If $U = 2^k$, then the distribution simplifies to $\pr(x) = 1/U$ (i.e., \emph{uniform}).

The following definition will be useful.
Consider $x \in [0,b-1]$ for some $b>0$ and let $c = \lceil \log_2 b \rceil$.
We define the \emph{minimal} binary codeword assigned to $x$ in the interval $[0,b-1]$
as $\B(x,c-1)$ if $x < 2^c - b$, $\B(x + 2^c - b, c)$ otherwise.
Note that if $b$ is a power of two the minimal binary codeword for $x$
is $\B(x,c)$.

\subsection{Gamma and Delta}\label{subsec:gamma_delta}

The two codes we now describe were introduced by~\citet*{elias75}
and are called \emph{universal} because the length of these codes
is $O(\log x)$ bits for every integer $x$, thus a constant factor away from
the optimal binary representation of length $|\B(x)| = \lceil \log_2(x+1) \rceil$ bits.
Additionally, they are uniquely-decodable.

The \emph{gamma} code for $x$ is
made by the unary representation of $|\B(x)|$
followed by the $|\B(x)| - 1$ least significant bits of $\B(x)$.
Therefore, $|\gamma(x)| = 2|\B(x)|-1$ bits
and $\pr(x) \approx 1/(2x^2)$.
Bit-by-bit decoding of $\gamma(x)$ is simple too.
First read the unary code, say $\ell$.
Then sum to
$2^{\ell-1}$
the integer represented by the next $\ell-1$ bits.
For example, the integer 113 is represented as
$\gamma(113) = \bit{1111110.110001}$,
because $\B(113) = \bit{1110001}$
is 7 bits long.

The key inefficiency of the gamma code lies in the use of the unary code
for the representation of $|\B(x)|$, which may become very large for big integers.
To overcome this limitation,
the \emph{delta} code replaces $\enc{U}{|\B(x)|}{}$ with $\gamma(|\B(x)|)$
in the $\delta$ representation of $x$.
The number of bits required by $\delta(x)$ is, therefore,
$|\gamma(|\B(x)|)| + |\B(x)| - 1$.
The corresponding distribution is $\pr(x) \approx 1/(2x(\log_2 x)^2)$.
Bit-by-bit decoding of $\delta$ codes follows automatically from that of $\gamma$ codes.
Again, the integer 113 is represented as $\delta(113) = \bit{11011.110001}$.
The first part of the encoding, \bit{11011}, is the $\gamma$ representation of 7, which is
the length of $\B(113)$.
Table~\ref{tab:codes} shows the integers 1..8 as encoded
with $\gamma$ and $\delta$ codes.

\begin{table}
\caption{The integers 1..8 as represented with several codes.
The ``.'' symbol highlights the distinction between different parts of the codes
and has a purely illustrative purpose:
it is not included in the final coded representation.
\label{tab:codes}}
\scalebox{\mytablescale}{\begin{tabular}{c lllllll}
\toprule

$x$ & $\enc{U}{x}{}$ & $\enc{B}{x}{}$ & $\gamma(x)$ & $\delta(x)$ & $\enc{G}{x}{2}$ & $\enc{ExpG}{x}{2}$ & $\enc{Z}{x}{2}$
\\

\midrule


1 & \bit{0}        & \bit{0}   &  \bit{0.}       & \bit{0.}        & \bit{0.0}    & \bit{0.00}  & \bit{0.0}     \\ 
2 & \bit{10}       & \bit{1}   &  \bit{10.0}     & \bit{100.0}     & \bit{0.1}    & \bit{0.01}  & \bit{0.10}    \\ 
3 & \bit{110}      & \bit{10}  &  \bit{10.1}     & \bit{100.1}     & \bit{10.0}   & \bit{0.10}  & \bit{0.11}    \\ 
4 & \bit{1110}     & \bit{11}  &  \bit{110.00}   & \bit{101.00}    & \bit{10.1}   & \bit{0.11} & \bit{10.000}  \\ 
5 & \bit{11110}    & \bit{100} &  \bit{110.01}   & \bit{101.01}    & \bit{110.0}  & \bit{10.000} & \bit{10.001}  \\ 
6 & \bit{111110}   & \bit{101} &  \bit{110.10}   & \bit{101.10}    & \bit{110.1}  & \bit{10.001} & \bit{10.010}  \\ 
7 & \bit{1111110}  & \bit{110} &  \bit{110.11}   & \bit{101.11}    & \bit{1110.0} & \bit{10.010} & \bit{10.011}  \\ 
8 & \bit{11111110} & \bit{111} &  \bit{1110.000} & \bit{11000.000} & \bit{1110.1} & \bit{10.011} & \bit{10.1000} \\ 

\bottomrule
\end{tabular}
}
\end{table}


In order to decode gamma codes faster on modern processors, a simple variant of gamma
is proposed by~\citet{schlegel2010fast} and called \emph{$k$-gamma}.
Groups of $k$ integers are encoded together, with $k = 2,3$ or 4, using the same number of bits.
Thus, instead of recording the unary length for each integer, only the length
of the largest integer in the group is written. This leads to a higher compression ratio
if the $k$ integers are close to each other, namely they require the same codeword length.
On the other hand, if the largest integers in the group requires more bits than the other
integers, this encoding is wasteful compared to the traditional gamma.
However, decoding is faster: once the binary length has been read, a group of $k$ integers
is decoded in parallel using SIMD instructions.
Similarly, ~\citet*{trotman2018elias} introduced a SIMD version of
delta codes.
A 512-bit payload is broken down into its 16 $\times$ 32-bit integers and the base-2 magnitude
of the largest integer is written using gamma coding as a selector
(writing the selector in unary code gives $16$-gamma).
Although not as fast as $k$-gamma, the representation is faster to decode compared to
decoding bit by bit.

\subsection{Golomb}

In 1966 Golomb introduced a parametric code
that is a hybrid between unary and binary~\cite{Golomb66}.
The Golomb code of $x$ with parameter $b > 1$,
$\enc{G}{x}{b}$,
consists in the representation of two pieces:
the \emph{quotient} $q = \lfloor (x-1)/b \rfloor$
and the \emph{remainder} $r = x - q \times b - 1$.
The quantity $q+1$ is written in unary,
followed by a minimal binary codeword assigned to $r \in [0,b-1]$.
Clearly, the closer $b$ is to the value of $x$ the smaller the value of $q$,
with consequent better compression and faster decoding speed.
Table~\ref{tab:codes} shows an example code with $b=2$.
Let us consider the code with $b=5$, instead.
From the definition of minimal binary codeword,
we have that $c = \lceil \log_2 5 \rceil = 3$ and $2^c - b = 3$.
Thus the first 3 reminders, 0..2, are always assigned the first 2-bit codewords
\bit{00}, \bit{01} and \bit{10} respectively.
The reminders 3 and 4 are instead assigned codewords \bit{110} and \bit{111}
as given by $\B(3 + 3, 3)$ and $\B(4 + 3, 3)$, respectively.
Decoding just reverts the encoding procedure.
After the unary prefix, always read $c-1$ bits, with $c = \lceil \log_2 b \rceil$.
If these $c-1$ bits give a quantity $r$ that is less than $2^c-b$, then stop
and work with the reminder $r$.
Instead, if $r \geq 2^c-b$ then fetch another bit and 
compute $r$ as the difference between this $c$-bit number and the quantity $2^c-b$.

Golomb was the first to observe that if $n$ integers are drawn \emph{at random}
from a universe of size $U$, then the
gaps between the integers follow a geometric distribution
$\pr(x) = p(1-p)^{x-1}$ with parameter
$p = n/U$ being the probability to find an integer $x$ among the ones selected.
It is now clear that
the optimal value for $b$ depends on $p$ and it
can be shown that this value is the integer closest
to $-1/\log_2(1-p)$, i.e., the value that
satisfies $(1-p)^b \approx 1/2$.
Doing the math we have
$b \approx 0.69 / p$,
which is a good approximation of the optimal value
and can be used to define a Golomb code with parameter $b$.
This code is optimal for the geometric distribution $\pr(x) = p(1-p)^{x-1}$.
~\citet*{gallager1975optimal} showed that the optimal
value for $b$ can be computed as
$b = - \lceil \log_2(2-p)/\log_2(1-p) \rceil$.


\subsection{Rice}\label{sec:rice}

The Rice code~\cite{Rice71,Rice91} is actually a special case of the Golomb code
for which $b$ is set to $2^k$, for some $k > 0$ (sometimes also referred
to as the Golomb-Rice code).
Let $\rice_k(x)$ be the Rice code of $x$ with parameter $k > 0$.
In this case the remainder $r$ is always written in $k$ bits.
Therefore, the length of the Rice code is
$|\rice_k(x)| = \lfloor (x-1)/2^k \rfloor + k + 1$ bits.
To compute the optimal parameter for the Rice code,
we just pretend to be constructing an optimal Golomb code
with parameter $b$ and then find two integers, $l$ and $r$, such that
$2^l \leq b \leq 2^r$. One of these two integers will be
the optimal value of $k$ for the Rice code.
(We also point the interested reader to the
technical report by~\citet*{kiely2004selecting} for
a deep analysis about the optimal values of the Rice parameter.)

\subsection{Exponential Golomb}

The \emph{exponential} Golomb code proposed by~\citet{teuhola1978compression}
logically defines a vector of ``buckets''
$$B = \Big[0, 2^k, \sum_{i=0}^1 2^{k+i}, \sum_{i=0}^2 2^{k+i}, \sum_{i=0}^3 2^{k+i}, \ldots\Big],\text{ for some } k \geq 0$$
and encodes an integer $x$ as a bucket identifier plus an offset relative to the bucket.
More specifically, the code $\enc{ExpG}{x}{k}$ is obtained as follows.
We first determine the bucket where $x$ belongs to, i.e., the index $h \geq 1$ such that
$B[h] < x \leq B[h+1]$. Then $h$ is coded in unary, followed by a minimal binary
codeword assigned to $x-B[h]-1$ in the shrunk interval $[0, B[h+1]-B[h]-1]$.
(Since $\log_2(B[h+1]-B[h])$ is always a power of 2 for the choice of $B$ above,
the binary codeword of $x$ is $\B(x-B[h]-1,\log_2(B[h+1]-B[h]))$.)

Table~\ref{tab:codes} shows an example for $k=2$.
Note that $\textsf{ExpG}_0$ coincides with Elias' $\gamma$.

\subsection{Zeta}

\citet{boldi2004webgraph,2005:boldi-vigna} introduced the family of \emph{zeta} codes
that is optimal for integers distributed according to a power law
with small exponent $\alpha$ (e.g., less than 1.6),
that is $\mathbb{P}(x) = 1/(\zeta(\alpha) x^\alpha)$, where $\zeta(\cdot)$ denotes the Riemann
zeta function.
The zeta code $\textsf{Z}_k$ is an exponential Golomb code relative to a vector of
``buckets'' $[0, 2^k-1, 2^{2k}-1, 2^{3k}-1, \ldots]$.
Again, Table~\ref{tab:codes} shows an examples for $k=2$.
Note that $\textsf{Z}_1$ coincides with $\textsf{ExpG}_0$, therefore
also $\textsf{Z}_1$ is identical to Elias' $\gamma$.

For example, let us consider $\enc{Z}{5}{2}$.
The value of $h$ is 2 because $2^2-1 < 5 < 2^4-1$. Therefore the first part
of the code is the unary representation of 2.
Now we have to assign a minimal binary codeword to $5 - (2^2-1) - 1 = 1$
using 3 bits, that is $\B(1,3) = \bit{001}$.
A more involved example is the one for, say, $\enc{Z}{147}{3}$.
In this case, we have $h=3$, thus the interval of interest is $[2^6-1,2^9-1]$.
Now we have to assign $147-(2^6-1)-1=83$ a minimal binary codeword in the interval $[0,448]$.
Since $83$ is more than the left extreme $2^6-1$, we have to write $\B(147,9)$
for a final codeword of \bit{110.010010011}.

\begin{table}
\caption{The integers 1..8 as represented with Fibonacci-based codes.
In (a), the final control bit is highlighted in bold font and
the relevant Fibonacci numbers $F_i$ involved in the representation
are also shown at the bottom of the table.
In (b), the ``canonical'' lexicographic codewords are presented.
\label{tab:fibonacci}}
\subfloat[``original'' codewords]{\scalebox{\mytablescale}{\begin{tabular}{ccccccc}
\toprule

$x$ & \multicolumn{6}{l}{$\enc{F}{x}{}$} \\

\midrule


1 & \bit{1} & \textbf{\bit{1}} & & & & \\
2 & \bit{0} & \bit{1} & \textbf{\bit{1}} & & & \\
3 & \bit{0} & \bit{0} & \bit{1} & \textbf{\bit{1}} & & \\
4 & \bit{1} & \bit{0} & \bit{1} & \textbf{\bit{1}} & & \\
5 & \bit{0} & \bit{0} & \bit{0} & \bit{1} & \textbf{\bit{1}} & \\
6 & \bit{1} & \bit{0} & \bit{0} & \bit{1} & \textbf{\bit{1}} & \\
7 & \bit{0} & \bit{1} & \bit{0} & \bit{1} & \textbf{\bit{1}} & \\
8 & \bit{0} & \bit{0} & \bit{0} & \bit{0} & \bit{1} & \textbf{\bit{1}} \\

\midrule
$F_i$ & 1 & 2 & 3 & 5 & 8 & 13 \\

\bottomrule
\end{tabular}
}
\label{tab:fibonacci_original}}
\hspace{1cm}
\subfloat[lexicographic codewords]{\scalebox{\mytablescale}{\begin{tabular}{ccccccc}
\toprule

$x$ & \multicolumn{6}{l}{$\enc{F}{x}{}$} \\

\midrule


1 & \bit{0} & \bit{0} & & & & \\
2 & \bit{0} & \bit{1} & \bit{0} & & & \\
3 & \bit{0} & \bit{1} & \bit{1} & \bit{0} & & \\
4 & \bit{0} & \bit{1} & \bit{1} & \bit{1} & & \\
5 & \bit{1} & \bit{0} & \bit{0} & \bit{0} & \bit{0} & \\
6 & \bit{1} & \bit{0} & \bit{0} & \bit{0} & \bit{1} & \\
7 & \bit{1} & \bit{0} & \bit{0} & \bit{1} & \bit{0} & \\
8 & \bit{1} & \bit{0} & \bit{0} & \bit{1} & \bit{1} & \bit{0} \\

\bottomrule
\end{tabular}
}
\label{tab:fibonacci_lex}}
\end{table}

\subsection{Fibonacci}

\citet*{fraenkel1985robust} introduced in 1985 a class of codes based on \emph{Fibonacci} numbers~\cite{Fibonacci}
and later generalized by~\citet*{apostolico1987robust}.
The encoding is a direct consequence of the \emph{Zeckendorf's theorem}: every positive integer
can be uniquely represented as the sum of some, non adjacent, Fibonacci numbers.
Let $F_i = F_{i-1} + F_{i-2}$ define the $i$-th Fibonacci number for $i > 2$, with $F_1 = 1$ and $F_2 = 2$.
Then we have: $F_3 = 3$, $F_4 = 5$, $F_5 = 8$, $F_6 = 13$, etc.
The Fibonacci encoding $\enc{F}{x}{}$ of an integer $x$ is obtained by:
(1) emitting a \bit{1} bit if the $i$-th Fibonacci number is used in the sum giving $x$, or emitting a \bit{0} bit otherwise; (2) appending a final control \bit{1} bit to ensure unique decodability.
Table~\ref{tab:fibonacci_original} shows the first 8 integers
as encoded with this procedure, where we highlighted in bold font
the final control bit.
For example, $7 = F_2 + F_4$, thus $\enc{F}{7}{}$ will be given by 4 bits where the second and the fourth
are \bit{1}, i.e., \bit{0101}, plus the control \bit{1} bot for a final codeword of \bit{01011}.

Note that the codewords assigned by the procedure described above
are not lexicographically-sorted in the integers they represent.
However, if we first compute the codeword lengths we can then generate
a set of lexicographically-sorted codewords in a rather simple way,
therefore obtaining a Fibonacci-based code that can be encoded/decoded
with the procedures we have illustrated in Section~\ref{sec:encoding_decoding}.
Given a non-decreasing sequence of codeword lengths $[\ell_1,\ldots,\ell_n]$
satisfying the Kraft-McMillan inequality (see the beginning of Section~\ref{sec:integer}),
the corresponding codewords are generated as follows.
The first codeword is always the bit string of length $\ell_1$ that is $\bit{0}^{\ell_1}$.
Now, let $\ell = \ell_1$.
For all $i = 2,\ldots,n$ we repeat the following two steps.
\begin{enumerate}
\item Let $c$ be the next lexicographic codeword of $\ell$ bits.
If $\ell_i = \ell$, then we just emit $c$.
Otherwise, $c$ is
padded with possible \bit{0} bits to the right until we have a $\ell_i$-bit codeword.
\item We set $\ell = \ell_i$.
\end{enumerate}
Note that the way be define $c$ in step (1) guarantees
that the generated code is prefix-free.

For our example in Table~\ref{tab:fibonacci_lex}, the sequence of codeword
lengths is $[2,3,4,4,5,5,5,6]$.
Let us generate the first 4 codewords.
The first codeword is therefore \bit{00}, with $\ell_1 = 2$.
The next codeword length $\ell_2$ is 3, thus we 
pad the next 2-bit codeword following \bit{00}, i.e., \bit{01},
with a \bit{0} and obtain the 3-bit codeword \bit{010}.
The next codeword length is 4, thus we obtain the codeword \bit{0110}.
The next codeword length is 4 again and the codeword is just obtained by assigning the
codeword following \bit{0110} in lexicographic order, that is \bit{0111}.

There is a closed-form formula for computing the $i$-th Fibonacci number, $i \geq 1$, called Binet's formula:
$$
	F_i = \frac{1}{\sqrt{5}} \Big[\Big(\frac{1+\sqrt{5}}{2}\Big)^{i+1} - \Big(\frac{1-\sqrt{5}}{2}\Big)^{i+1}\Big] \approx \Big(\frac{1 + \sqrt{5}}{2}\Big)^{i+1} = \phi^{i+1},
$$
where $\phi = \frac{1 + \sqrt{5}}{2}$ is the so-called \emph{golden ratio}.
Using this formula, it can be shown that the codeword length of $\enc{F}{x}{}$
is approximately equal to $1 + \log_{\phi} x$ bits.
Therefore, the corresponding distribution is
$\pr(x) = 1/(2x^{1/\log_2\phi}) \approx 1/(2x^{1.44})$.
This implies that Fibonacci-based codes are shorter than $\gamma$ for $x>3$;
and as good as or even better than $\delta$ for a wide range of practical values
ranging from $F_2=2$ to $F_{19}=\num{6765}$.

\subsection{Variable-Byte}\label{subsec:vb}

The codes described in the previous sections are \emph{bit-aligned}
as they do not represent an integer using a multiple of a fixed number of bits,
e.g., a byte.
But reading a stream of bits in chunks where each chunk is
a byte of memory (or a multiple of a byte, e.g., a memory word -- 4 or 8 bytes),
is simpler and faster because
the data itself is written in memory in this way.
Therefore, it could be preferable to use
\emph{byte-aligned} or \emph{word-aligned} codes
when decoding speed is the main concern rather than compression effectiveness.

\emph{Variable-Byte}, first described by~\citet*{thiel1972program}, is the most popular and simplest byte-aligned code: the binary representation of a non-negative integer is split into groups of $7$ bits which are represented as a sequence of bytes. In particular, the $7$ least significant bits of each byte are reserved for the data whereas the most significant, called the \emph{continuation bit}, is equal to $1$ to signal continuation of the byte sequence. The last byte of the sequence has its $8$-th bit set to $0$ to signal, instead, the termination of the byte sequence.
The main advantage of Variable-Byte codes is decoding speed: we just need to read one byte at a time until we found a value smaller than $2^7$.
Conversely, the number of bits to encode an integer cannot be less than $8$, thus Variable-Byte is only suitable for large numbers and its compression ratio may not be competitive with the one of bit-aligned codes for small integers.
Variable-Byte uses $\lceil \lceil \log_2(x + 1) \rceil / 7 \rceil \times 8$
bits to represent the integer $x$, thus
it is optimal for the distribution $\pr(x) \approx \sqrt[7]{1/x^8}$.
For example, the integer \num{65790} is represented as
\bit{\textbf{0}0000100.\textbf{1}0000001.\textbf{1}1111110},
where we mark the control bits in bold font. Also notice the padding bits in the first byte starting from the left, inserted to align the binary representation of the number to a multiple of $8$ bits.

\emph{Nibble} coding is a simple variation of this strategy where 3 bits are used for data
instead of 7,
which is optimal for the distribution $\pr(x) \approx \sqrt[3]{1/x^4}$.

\citet*{culpepper2005enhanced} describe a byte-aligned
code with the property that the first byte of each
codeword defines the length of the codeword, which
makes decoding simpler and faster.

Various enhancements were proposed to accelerate the sequential decoding speed of Variable-Byte.
For example, in order to reduce the probability of a branch misprediction that leads to higher throughput and helps keeping the CPU pipeline fed with useful instructions, the control bits can be grouped together.
If we assume that the largest represented integer fits into four bytes, we have to distinguish between only four different byte-lengths, thus two bits are sufficient.
In this way, groups of four integers require one control byte only.
This optimization was introduced in Google's {\varigb} format~\cite{2009:dean},
which is faster to decode than the original Variable-Byte code.

Working with byte-aligned codes also opens the possibility of exploiting the parallelism of SIMD instructions to further enhance the sequential decoding speed.
This is the approach taken by the proposals {\varigiu}~\cite{2011:stepanov.gangolli.ea}, {\masked}~\cite{2015:plaisance.kurz.ea} and {\stream}~\cite{2018:lemire.kurz.ea} that we overview below.

{\varigiu}~\cite{2011:stepanov.gangolli.ea} uses a format similar to the one of {\varigb}: one control byte describes a variable number of integers in a data segment of exactly eight bytes, therefore each group can contain between two and eight compressed integers.
{\masked}~\cite{2015:plaisance.kurz.ea} works, instead, directly on the original Variable-Byte format. The decoder first gathers the most significant bits of consecutive bytes using a dedicated SIMD instruction. Then using previously-built look-up tables and a shuffle instruction, the data bytes are permuted to obtain the decoded integers.
{\stream}~\cite{2018:lemire.kurz.ea} separates the encoding of the control bits from the data bits by writing them into separate streams. This organization permits to decode multiple control bits simultaneously and, consequently, to reduce data dependencies
that can stop the CPU pipeline execution when decoding the data stream.

\begin{table}
\caption{The integers 1..20 as represented by $\enc{SC}{4,4}{}$- and $\enc{SC}{5,3}{}$-dense codes respectively.
\label{tab:sc}}

    \scalebox{\mytablescale}{
    	\begin{tabular}{cll}
\toprule

$x$ & $\enc{SC}{4,4,x}{}$ & $\enc{SC}{5,3,x}{}$ \\

\midrule


1  & \bit{000} & \bit{000} \\
2  & \bit{001} & \bit{001} \\
3  & \bit{010} & \bit{010} \\
4  & \bit{011} & \bit{011} \\
5  & \bit{100.000} & \bit{100} \\
6  & \bit{100.001} & \bit{101.000} \\
7  & \bit{100.010} & \bit{101.001} \\
8  & \bit{100.011} & \bit{101.010} \\
9  & \bit{101.000} & \bit{101.011} \\
10 & \bit{101.001} & \bit{101.100} \\

\bottomrule
\end{tabular}

    }
\hspace{0.5cm}
    \scalebox{\mytablescale}{
    	\begin{tabular}{cll}
\toprule

$x$ & $\enc{SC}{4,4,x}{}$ & $\enc{SC}{5,3,x}{}$ \\

\midrule


11 & \bit{101.010} & \bit{110.000} \\
12 & \bit{101.011} & \bit{110.001} \\
13 & \bit{110.000} & \bit{110.010} \\
14 & \bit{110.001} & \bit{110.011} \\
15 & \bit{110.010} & \bit{110.100} \\
16 & \bit{110.011} & \bit{111.000} \\
17 & \bit{111.000} & \bit{111.001} \\
18 & \bit{111.001} & \bit{111.010} \\
19 & \bit{111.010} & \bit{111.011} \\
20 & \bit{111.011} & \bit{111.100} \\

\bottomrule
\end{tabular}

    }
\end{table}

\subsection{SC-Dense}

In Variable-Byte encoding the value $2^7$ acts as a separator between \emph{stoppers},
i.e., all values in $[0, 2^7-1]$, and \emph{continuers}, i.e., all values in $[2^7, 2^8-1]$.
A generalization of the encoding can be obtained by changing the separator value,
thus enlarging or restricting the cardinalities of the set of continuers and stoppers.
In general, the values from 0 to $c-1$ are reserved to stoppers and the values from $c$
to $c+s-1$ to continuers, provided that $c+s=2^8$.
Intuitively, changing the separating value can better adapt to the distribution of the integers
to be encoded. For example, if most integers are larger than (say) 127, then it is
convenient to have more continuers.
This is the main idea behind the \textsf{SC}-dense code introduced by~\citet{brisaboa2003s}.

Given the integer $x$, its $\enc{SC}{s,c,x}{}$ representation is obtained as follows.
Let $k(x) \geq 1$ be the number of $\lceil \log_2 (s+c) \rceil$-bit words
needed by the representation of $x$.
This value $k(x)$ will be such that
$$s\frac{c^{k(x)-1}-1}{c-1} \leq x < s\frac{c^{k(x)}-1}{c-1}.$$
If $k(x)=1$, the representation is just the stopper $x-1$.
Otherwise, let $y = \lfloor (x-1)/s \rfloor$ and $x^\prime = x - (sc^{k(x)-1} - s)/(c-1)$.
In this case, $x > s$ and $k(x)$ is given by $\lfloor (y-1)/c \rfloor + 2$.
The representation is given by a sequence of $k(x)-1$ continuers,
made by repeating the continuer $s+c-1$ for $k(x)-2$ times
followed by the continuer $s+((y-1)$ mod $c)$,
plus the final stopper $(x^\prime - 1)$ mod $s$.
The number of bits required by encoding of $x$ is $k(x) \lceil \log_2(s+c) \rceil$,
thus it follows that $\pr(x) \approx (s+c)^{-k(x)}$.
It is also possible to compute via dynamic programming the optimal values
for $s$ and $c$ given the probability distribution of the integers~\cite{brisaboa2003s}.

Table~\ref{tab:sc} shows the codewords assigned to the integers 1..20 by the dense codes
$\enc{SC}{4,4}{}$ and $\enc{SC}{5,3}{}$ respectively.
Let us consider the encoding of $x=13$ under the code $\enc{SC}{5,3}{}$.
In this example, we have $y = \lfloor (13-1)/5 \rfloor = 2$ and $k(13) = \lfloor (2-1)/3 \rfloor + 2 = 2$.
The only continuer is therefore given by $5 + ((2-1)$ mod $3) = 6$, i.e., \bit{110}.
Since $x^\prime = 3$, the stopper is $(3-1)$ mod $3=2$, i.e., \bit{010},
for a final representation of \bit{110.010}.


\begin{figure}
\centering
\includegraphics[scale=0.73]{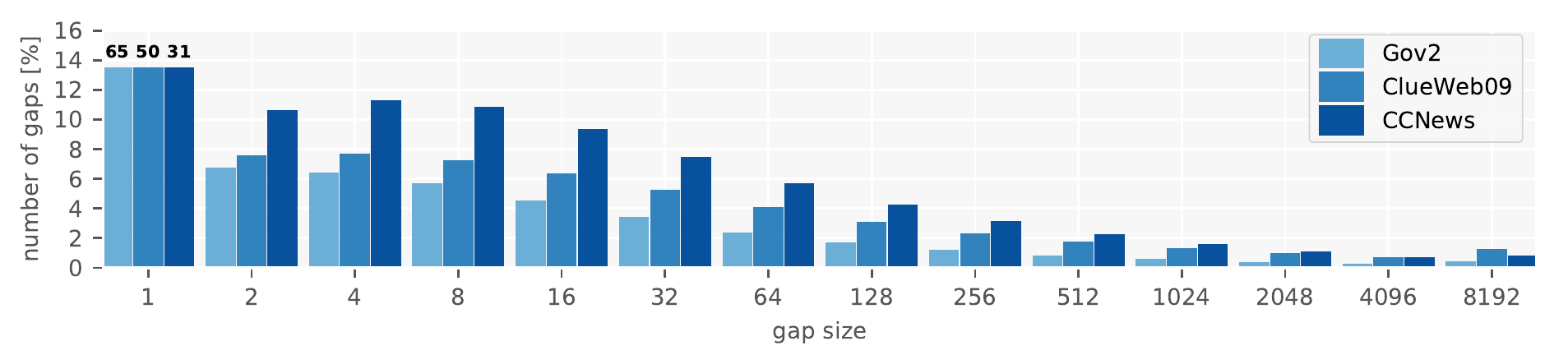}
\caption{Distribution of the gaps in the real-world datasets {\gov}, {\clue}, and {\cc}.
\label{fig:dgaps_distribution}}
\end{figure}

\subsection{Concluding remarks}

In the context of inverted indexes, we can exploit the fact that
inverted lists are sorted -- and typically, strictly increasing --
to achieve better compression.
In particular, given a sequence $\seq[1..n]$ of this form,
we can transform the sequence into $\seq^{\prime}$
where $\seq^{\prime}[i] = \seq[i] - \seq[i-1]$
for $i > 1$ and $\seq^{\prime}[1] = \seq[1]$.
In the literature, $\seq^{\prime}$ is said to be formed by the so-called
\emph{gaps} of $\seq$ (or \emph{delta-gaps}).
Using the codes described in this section on the gaps of $\seq$ is a very
popular strategy for compressing inverted indexes,
with the key requirement of performing a prefix-sum during decoding.
Clearly, compressing these gaps is far more convenient than compressing the
integers of $\seq$ because the gaps are smaller, thus
less bits are required for their codes.
In this case, the compressibility of the inverted index critically depends
on the distribution of the gaps but, fortunately, most of them are small.
Fig.~\ref{fig:dgaps_distribution} shows the distribution of the gaps
for three large text collection that we will introduce in Section~\ref{sec:experiments}
(see Table~\ref{tab:datasets}, at page~\pageref{tab:datasets}, for their basic statistics).
The plot highlights the \emph{skewed} distribution of the gaps: the most frequent
integer is a gap of 1 so that, for better visualization, we cut the percentage to 16\%
but report the actual value in bold.
For example, on the {\clue} dataset 50\% of the
gaps are just 1.
The other values have decreasing frequencies.
We divide the distribution into buckets of exponential size,
namely the buckets $B = [1,2,4,8,\ldots,8192]$.
In particular, bucket $B[i]$
comprises all gaps $g$ such that $B[i-1] < g \leq B[i]$.
(The last bucket also comprises all other gaps larger than $8192$
-- the ``tail'' of the distribution.)

\begin{figure}
\centering
	\includegraphics[scale=0.73]{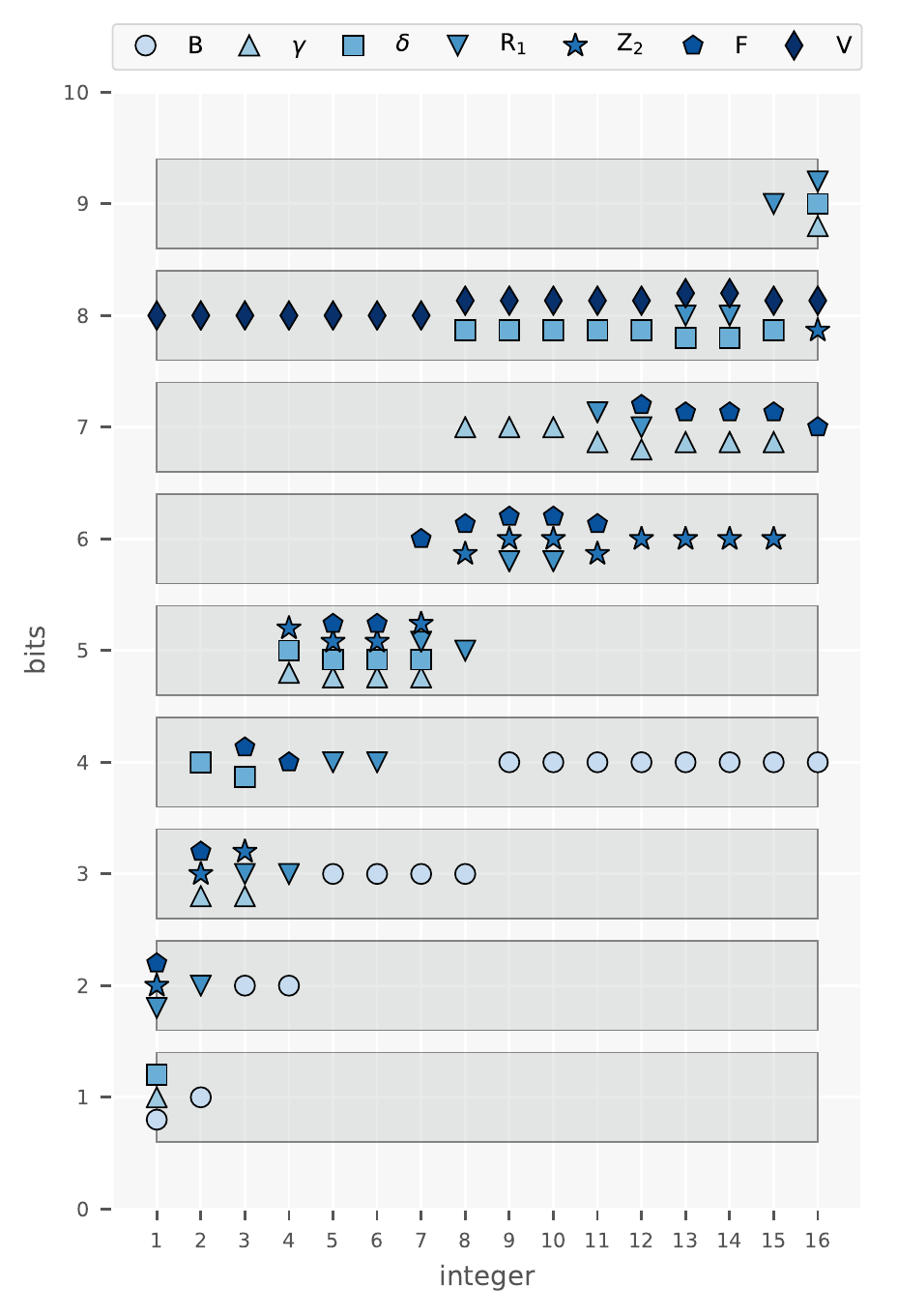}
	\label{fig:codes_comparison_a}
	\includegraphics[scale=0.73]{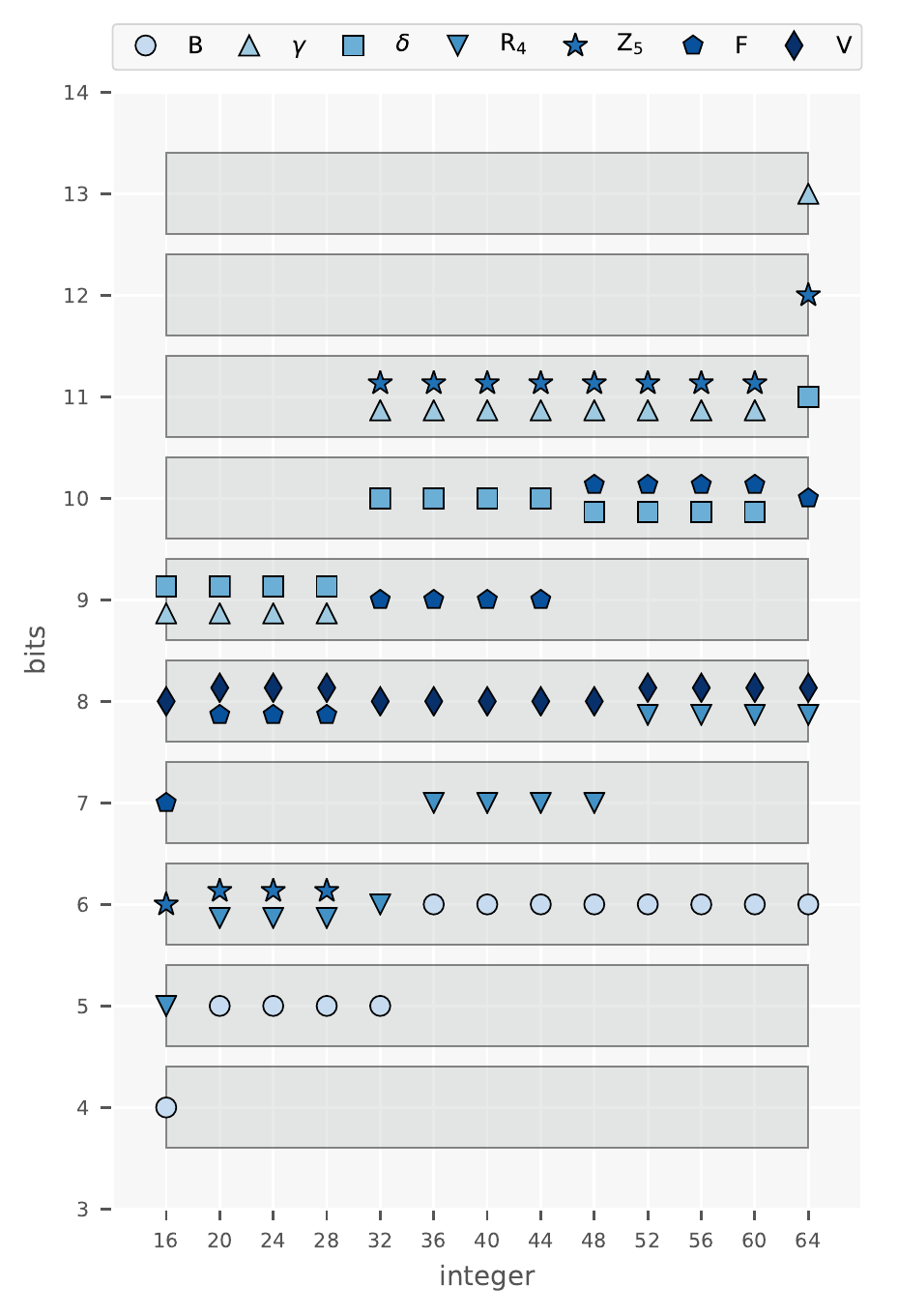}
	\label{fig:codes_comparison_b}
\caption{Comparison between several codes described in Section~\ref{sec:integer} for the integers 1..64.
\label{fig:codes_comparison}}
\end{figure}

As an illustrative comparison between several of the codes described in this section,
we report in Fig.~\ref{fig:codes_comparison} the number of bits taken
by their codewords when representing the integers 1..64,
knowing that such values cover most of the gaps we have in inverted index data
(e.g., approximately 86 -- 95\% of the gaps shown in Fig.~\ref{fig:dgaps_distribution}).
In particular, we show the comparison between the codes:
binary (\textsf{B})
as an illustrative ``single-value'' lower bound,
$\gamma$, $\delta$,
Rice (\textsf{R}), Zeta (\textsf{Z}), Fibonacci (\textsf{F}) and
Variable-Byte (\textsf{V}).
In the plots, data points corresponding to different methods
that have the same coordinates have been stacked vertically
for better visualization otherwise these would have been indistinguishable.
For example,
$\gamma$, $\delta$, Zeta and Fibonacci all take 5 bits to represent
the integers 5 and 6.
Not surprisingly, a tuned parametric code such as Rice or Zeta may be the best choice.
However, tuning is not always possible and a single parameter has to be specified for
the encoding of all integers in the list (or, say, in a sufficiently large block).
For the smallest integers
the universal codes $\gamma$ and $\delta$ are very good, but not competitive
immediately for slightly larger integers, e.g., larger than 16.
On such values and larger, a simple byte-aligned
strategy such as Variable-Byte performs well.

\section{List Compressors}\label{sec:list}

In this section
we describe algorithms that encode an integer list,
instead of representing each single integer separately.
A useful tool to analyze the space effectiveness of these compressors
is the combinatorial \emph{information-theoretic} lower bound,
giving the minimum number of bits needed to represent a list of $n$
\emph{strictly increasing} integers drawn \emph{at random}
from a universe of size $U \geq n$, that is~\cite{pagh2001low}
($e = 2.718$ is the base of the natural logarithm):
$$\Big\lceil \log_2 {U \choose n} \Big\rceil 
= n \log_2(eU/n) - \Theta(n^2/U) - O(\log n)$$
which is approximately
$$n(\log_2(U/n) + \log_2 e) = 
n\log_2(U/n) + 1.443n
\text{ bits, for } n = o(\sqrt{U}).$$

However, it is important to keep in mind
that the compressors we describe in this section
often take less space than that computed using
the information-theoretic lower bound.
In fact, while the lower bound assumes that the integers
are distributed at random,
the compressors
take advantage of the fact that inverted lists feature
\emph{clusters} of close integers,
e.g., runs of consecutive integers, that are far more compressible than highly scattered regions.
This is also the reason why these compressors usually outperform
the ones presented in Section~\ref{sec:integer}, at least for sufficiently long lists.
As a preliminary example of exploitation of such local clusters,
we mention the \emph{Binary Adaptive Sequential Coding} (\textsf{BASC}) by~\citet*{moffat2006binary}.
Given a sequence of integers $\seq[1..n]$,
instead of coding the binary magnitude $b_i = \lceil \log_2(\seq[i] + 1) \rceil$
of every single integer $\seq[i]$ -- as it happens, for example,
in the Elias' and related codes --
we can assume $b_{i+1}$ to be similar to $b_i$ (if not the same).
This allows to code $b_{i+1}$ \emph{relatively} to $b_i$, hence amortizing its cost.

Such natural clusters of integers are present because the indexed documents themselves
tend to be clustered, i.e., there are subsets of documents sharing the very same set of terms.
Consider all the Web pages belonging to a certain domain:
since their topic is likely to be the same, they are also likely to share a lot of terms.
Therefore, not surprisingly, list compressors greatly benefit from {\docid}-reordering strategies
that focus on re-assigning the {\docid}s in order to form larger clusters.
When the indexed documents are Web pages, a simple and effective strategy is to assign identifiers to documents
according to the lexicographic order of their URLs~\cite{2007:silvestri}.
This is the strategy we use in the experimental analysis in Section~\ref{sec:experiments}
and its benefit is highlighted by Fig.~\ref{fig:dgaps_distribution}: the most frequent
gap size is just 1.
Another approach uses a recursive graph bisection algorithm to find a suitable
re-ordering of {\docid}s~\cite{2016:dhulipala.kabiljo.ea}.
In this model, the input graph is a bipartite graph in which one set of vertices
represents the terms of the index and the other set represents the {\docid}s.
A graph bisection identifies a permutation of the {\docid}s and, thus, the goal is that of finding,
at each step of recursion,
the bisection of the graph which minimizes the size of the graph compressed using delta encoding.
(There are also other reordering strategies that may be relevant~\cite{blandford2002index,shieh2003inverted}:
the works cited here are not meant to be part of an exhaustive list.)

\subsection{Binary packing}

A simple way to improve both compression ratio and decoding speed
is to encode a \emph{block} of integers, instead of the whole sequence.
This line of work finds its origin in the so-called
\emph{frame-of-reference} (\textsf{FOR})~\cite{GoldsteinRS98}.
Once the sequence has been partitioned into blocks (of fixed or variable length),
then each block is encoded separately.
The key insight behind this simple idea is that, if the sequence is locally
homogeneous, i.e., it features clusters of close integers,
the values in a block are likely to be of similar magnitude.
Vice versa, it is generally hard to expect a long sequence to be globally
homogeneous and this is why compression on a per-block basis gives usually
better results.

An example of this approach is \emph{binary packing}.
Given a block, we can compute the bit width $b = \lceil\log_2(\var{max} + 1)\rceil$
of the \var{max} element in the block
and then represent all integers in the block using $b$-bit codewords.
Clearly the bit width $b$ must be stored prior to the representation of the block.
Moreover, the gaps between the integers can be computed to lower the value of $b$.
Many variants of this simple approach have been proposed~\cite{2010:silvestri.venturini,2012:delbru.campinas.ea,2013:lemire.boytsov}.
For example, in the \emph{Recursive Bottom-Up Coding} (\textsf{RBUC}) code proposed
by~\citet*{moffat2005binary}, blocks of fixed size are considered, the bit width $b_i$
of each block determined, and the $i$-th block represented via $b_i$-bit codewords.
The sequence of bit widths $\{b_i\}_i$ needs to be represented as well and the
procedure sketched above is applied recursively to it.

Dividing a list into fixed-size blocks may be suboptimal because
regions of close identifiers may be contained in a block containing a much larger value.
Thus, it would be preferable to split the list into \emph{variable}-size blocks
in order to better adapt to the distribution of the integers in the list.
~\citet*{2010:silvestri.venturini} present an algorithm
that finds the optimal splitting of a list of size $n$ in time $O(kn)$,
where $k$ is the maximum block size allowed, in order to minimize the overall encoding cost.
~\citet*{2013:lemire.boytsov} report that the
fastest implementation of this approach -- named \emph{Vector of Splits Encoding} (\textsf{VSE}) --
is that using splits of size 1..14, 16 and 32 integers.

~\citet*{2013:lemire.boytsov} propose word-aligned versions of binary packing for fast decoding.
In the scheme called {\bpt}, 4 groups of 32 bit-packed integers each are stored together in a meta block.
Each meta block is aligned to 32-bit boundaries and a 32-bit word is used as descriptor of the meta block.
The descriptor stores the 4 bit widths of the 4 blocks in the meta block (8-bit width for each block).
The variant called {\bpsimd} combines 16 blocks of 128 integers each that are aligned to
128-bit boundaries.
The use of SIMD instructions provides fast decoding speed.

\begin{table}
\caption{The 9 different ways of packing integers in a 28-bit segment as used by {\simplen}.
\label{tab:simple}}
\scalebox{\mytablescale}{\begin{tabular}{cccc}
\toprule

4-bit selector & integers & bits per integer & wasted bits  \\

\midrule

\bit{0000} & 28 & 1 & 0 \\
\bit{0001} & 14 & 2 & 0 \\
\bit{0010} & 9 & 3 & 1 \\
\bit{0011} & 7 & 4 & 0 \\
\bit{0100} & 5 & 5 & 3 \\
\bit{0101} & 4 & 7 & 0 \\
\bit{0110} & 3 & 9 & 1 \\
\bit{0111} & 2 & 14 & 0 \\
\bit{1000} & 1 & 28 & 0 \\

\bottomrule
\end{tabular}
}
\end{table}

\subsection{Simple}\label{subsec:simple}

Rather than splitting the sequence into blocks of integers as in binary packing,
we can split the sequence into fixed-memory units and ask how many integers can be
packed in a unit.
This is the key idea of the \emph{Simple} family of encoders introduced by~\citet*{AnhM05}:
pack as many integers as possible in a memory word, i.e., $32$ or $64$ bits.
This approach typically provides good compression and high decompression speed.

For example, {\simplen}~\cite{AnhM05} (sometimes also referred to as \textsf{Simple4b}~\cite{AnhM10})
adopts 32-bit memory words. It dedicates
4 bits to the \emph{selector} code and 28 bits for data.
The selector provides information about how many elements are packed in the data segment
using equally-sized codewords. A selector \bit{0000} may correspond to 28 1-bit integers;
\bit{0001} to 14 2-bit integers;
\bit{0010} to 9 3-bit integers (1 bit unused), and so on, as we can see
in Table~\ref{tab:simple}.
The four bits distinguish between 9 possible configurations.
Similarly, {\simples}~\cite{zlt08www} has 16 possible configurations using 32-bit words.
{\simplee}~\cite{AnhM10}, instead, uses 64-bit words with 4-bit selectors.
Dedicating 60 bits for data offers 14 different combinations rather than just 9,
with only 2 configurations having wasted bits rather than 3.

~\citet*{AnhM05} also describe two variations of the {\simplen} mechanism, named
\textsf{Relative10} and \textsf{Carryover12}.
The idea behind \textsf{Relative10} is to just use 2 bits for the selector, thus
allowing 10 packing configurations with 30 bits. In order to make use of more than
4 options, the selector code is combined with the one of the previous word, hence
enabling the whole range of 10 possibilities.
However, when packing 7 $\times$ 4-bit integers or 4 $\times$ 7-bit integers, two bits per word are wasted
(only 28 out of the 30 available bits are used).
Therefore, in the \textsf{Carryover12} approach these two bits are used to define
the selector code of the following word configuration that makes use of the full 32 bits
for packing the integers.

A similar approach to that of the Simple family is used
in the {\qmx} mechanism, introduced by~\citet*{2014:trotman}.
Considering memory words larger than 64 bits is a popular strategy for
exploiting the parallelism of SIMD instructions.
{\qmx} packs as many integers as possible into 128- or 256-bit words (Quantities)
and stores the selectors (eXtractors) separately in a different stream.
The selectors are compressed (Multipliers) with \emph{run-length encoding},
that is with a stream of pairs (\emph{value}, \emph{length}).
For example, given the sequence $[12,12,12,5,7,7,7,7,9,9]$,
its corresponding \textsf{RLE} representation is $[(12,3), (5,1), (7,4), (9,2)]$.

\subsection{PForDelta}\label{subsec:pfor}

The biggest limitation of block-based strategies is their space-inefficiency
whenever a block contains just one large value, because this forces the
compressor to use a universe of representation as large as that value.
This is the main motivation for the introduction of a
``patched'' frame-of-reference or
\emph{PForDelta} ({\pfor}), proposed by~\citet{zukowski06super}.
The idea is to choose a value $k$ for the universe of representation of the block, such that a large fraction, e.g., $90\%$, 
of its integers can be represented using $k$ bits per integer.
All integers that do not fit in $k$ bits, are treated as \emph{exceptions}
and encoded in a separate array using another compressor,
e.g., Variable-Byte or Simple.
This strategy is called \emph{patching}.
More precisely, two configurable parameters are chosen: a base value $b$ and a universe of representation $k$, so that most of the values fall in the range $[b, b + 2^k - 1]$ and can be encoded with $k$ bits each by shifting them (delta-encoding) in the range $[0, 2^k - 1]$. To mark the presence of an exception, we also need a special \emph{escape} symbol, thus we have $[0, 2^k-2]$ available configurations.

For example, the sequence $[$3, 4, 7, 21, 9, 12, 5, 16, 6, 2, 34$]$
is represented using PForDelta with parameters $b = 2$ and $k = 4$ as
$[1, 2, 5, \ast, 7, 10, 3, \ast, 4, 0, \ast] - [{21}, {16}, {34}]$.
The special symbol $\ast$ marks the presence of an exception
that is written in a separate sequence, here reported after the dash.

The \emph{optimized} variant {\opf} devised by~\citet*{2009:yan.ding.ea}, which selects for each block the values of $b$ and $k$ that minimize its space occupancy, it is more space-efficient and only slightly slower than the original {\pfor}.
~\citet*{2013:lemire.boytsov} proposed another variant called
\textsf{Fast-PFor} where exceptions are compressed
in \emph{pages},
i.e., groups of blocks of integers.
For example, a page may be 32 consecutive blocks
of 128 integers each, for a total of 4096 integers.
In this scheme, all the $b$-bit exceptions
from all the blocks in a page
are stored contiguously,
for $b = 1..32$.
What makes this organization faster to decode is
the fact that exceptions are
decoded in bulk at a page level, rather than at a (smaller)
block level as in {\opf}.

In the \emph{parallel} \textsf{PFor} method, proposed by~\citet{ao2011efficient},
exceptions are represented in a different way to allow
their decompression in parallel with that of the
``regular'' values.
Instead of using the escape symbol,
each time an exception $x$ is encountered
only
the least $k$ bits of $x-b$ 
are written
and the overflow bits accumulated in a separate array.
The positions of the exceptions are stored in another array and compressed
using a suitable mechanism.
The same sequence used in the above example, for $b=2$ and $k=4$,
becomes
$[1, 2, 5, \bold{3}, 7, 10, 3, \bold{14}, 4, 0, \bold{0}] - [1,0,2] - [4,8,11]$,
because: the least 4 bits of the exceptions $21-2=19$, $16-2=14$, and $34-2=32$
are 3, 14 and 0 respectively (in bold font); the corresponding overflow bits are 1, 0, and 2;
the three exceptions appear at positions 4, 8, and 11.

\subsection{Elias-Fano}\label{EF}

The encoder we now describe was independently proposed by~\citet*{Elias74} and~\citet*{Fano71}.
Let $\seq(n,U)$ indicate a sorted sequence $\seq[1..n]$
whose integers are drawn from a universe of size $U > \seq[n]$.
The binary representation of each integer $\seq[i]$ as $\B(\seq[i], \lceil\log_2 U \rceil)$
is split into two parts: a \emph{low} part consisting in the right-most $\ell = \lceil \log_2(U/n)\rceil$ bits that we call \emph{low bits} and a \emph{high} part consisting in the remaining $\lceil \log_2 U \rceil - \ell$ bits that we similarly call \emph{high bits}. Let us call $\ell_i$ and $h_i$ the values of low and high bits of $\seq[i]$ respectively.
The Elias-Fano encoding of $\seq(n,U)$ is given by the encoding of the high and low parts.
The integers $[\ell_1, \ldots, \ell_{n}]$ are written verbatim in
a bitvector $L$ of
$n \lceil \log_2(U/n)\rceil$ bits,
which represents the encoding of the low parts.
The high parts are represented with another bitvector
of $n + 2^{\lfloor \log_2 n \rfloor} \leq 2n$ bits as follows.
We start from a $0$-valued bitvector $H$ and set the bit in position $h_i + i$, for all $i = 1,\ldots,n$.
It is easy to see that the $k$-th unary value $m$ of $H$ indicates
that $m-1$ integers of $\seq$ have high bits equal to
$k$, $0 \leq k \leq \lfloor \log_2 n \rfloor$.
Finally the Elias-Fano representation of $\seq$ is given by the concatenation of $H$ and $L$ and overall takes
\begin{equation}\label{ef_space}
\ef(\seq(n,U)) \leq n\lceil\log_2(U/n)\rceil + 2n \mbox{ bits.}
\end{equation}
Although we can opt for an arbitrary split into high and low parts, ranging from $0$ to $\lceil \log_2 U \rceil$, it can be shown that $\ell = \lceil \log_2(U/n) \rceil$ minimizes the overall space occupancy of the encoding~\cite{Elias74}.
Moreover, given that the information-theoretic lower bound is approximately $n \log_2(U/n) + n\log_2 e$ bits,
it can be shown~\cite{Elias74} that less than half a bit is wasted per element by Formula~\ref{ef_space}.
Table~\ref{tab:elias_fano} shows an example of encoding for the sequence
$[$3, 4, 7, 13, 14, 15, 21, 25, 36, 38, 54, 62$]$.
Note that no integer has high part equal to \bit{101}.

\begin{table}
\caption{An example of Elias-Fano encoding applied to the sequence
$\seq = [3, 4, 7, 13, 14, 15, 21, 25, 36, 38, 54, 62]$.
\label{tab:elias_fano}}
\vspace{-0.5cm}
\scalebox{\mytablescale}{\begin{tabular}{l ccc|ccc|c|c|cc|c|c|c}
\toprule

$\seq$ & 3 & 4 & 7 & 13 & 14 & 15 & 21 & 25 & 36 & 38 &   & 54 & 62 \\

\midrule

\multirow{3}{*}{\emph{high}}

& \bit{0} & \bit{0} & \bit{0} & \bit{0} & \bit{0} & \bit{0} & \bit{0} & \bit{0} & \bit{1} & \bit{1} & \gray{\bit{1}} & \bit{1} & \bit{1} \\
& \bit{0} & \bit{0} & \bit{0} & \bit{0} & \bit{0} & \bit{0} & \bit{1} & \bit{1} & \bit{0} & \bit{0} & \gray{\bit{0}} & \bit{1} & \bit{1} \\
& \bit{0} & \bit{0} & \bit{0} & \bit{1} & \bit{1} & \bit{1} & \bit{0} & \bit{1} & \bit{0} & \bit{0} & \gray{\bit{1}} & \bit{0} & \bit{1} \\

\cmidrule(lr){1-14}

\multirow{3}{*}{\emph{low}}

& \bit{0} & \bit{1} & \bit{1} & \bit{1} & \bit{1} & \bit{1} & \bit{1} & \bit{0} & \bit{1} & \bit{1} &         & \bit{1} & \bit{1} \\
& \bit{1} & \bit{0} & \bit{1} & \bit{0} & \bit{1} & \bit{1} & \bit{0} & \bit{0} & \bit{0} & \bit{1} &         & \bit{1} & \bit{1} \\
& \bit{1} & \bit{0} & \bit{1} & \bit{1} & \bit{0} & \bit{1} & \bit{1} & \bit{1} & \bit{0} & \bit{0} &         & \bit{0} & \bit{0} \\

\cmidrule(lr){1-14}

$H$ &
\multicolumn{3}{c}{\bit{1110}} & \multicolumn{3}{c}{\bit{1110}} & \multicolumn{1}{c}{\bit{10}} &
\multicolumn{1}{c}{\bit{10}}   & \multicolumn{2}{c}{\bit{110}}  & \multicolumn{1}{c}{\bit{0}}  &
\multicolumn{1}{c}{\bit{10}}   & \multicolumn{1}{c}{\bit{10}} \\

\cmidrule(lr){1-14}

$L$ &
\multicolumn{3}{c}{\bit{011.100.111}} &
\multicolumn{3}{c}{\bit{101.110.111}} &
\multicolumn{1}{c}{\bit{101}} &
\multicolumn{1}{c}{\bit{001}} &
\multicolumn{2}{c}{\bit{100.110}} &
\multicolumn{1}{c}{} &
\multicolumn{1}{c}{\bit{110}} &
\multicolumn{1}{c}{\bit{110}} \\

\bottomrule
\end{tabular}
}
\end{table}

The same code arrangement was later described by~\citet*{Vo:1998:CIF:290941.291011}
as a ``modified'' version of the Rice code (Section~\ref{sec:rice}).
In fact, they partition $U$ into buckets of $2^k$ integers each
for some $k>0$,
code in unary how many integers fall in each bucket,
and represent each integer
using $k$ bits as an offset to its bucket.
The connection with Rice is established by writing the number of integers
sharing the same quotient, rather than encoding this quantity for every integer.
They also indicated that the optimal parameter $k$ should be chosen
to be $\lceil \log_2(U/n) \rceil$.

\parag{Supporting random Access}
Despite the elegance of the encoding, it is possible to support random access to individual
integers \emph{without} decompressing the whole sequence.
Formally, we are interested in implementing
the operation $\access(i)$ that returns $\seq[i]$.
The operation can be implemented by using an auxiliary data structure
that is built on the bitvector $H$ and efficiently answers $\select_1$ queries.
The answer to a $\select_b(i)$ query over a
bitvector is the
position of the $i$-th bit set to $b$.
For example, $\select_{\bit{0}}(3) = 10$ on the bitvector $H$ of Table~\ref{tab:elias_fano}.
This auxiliary data structure is \emph{succinct} in the sense that it is negligibly small in asymptotic terms, compared to
$\ef(\seq(n,U))$, requiring only $o(n)$ additional bits~\citep{MN07,2013:vigna}.
Using the $\select_{\bit{1}}$ primitive, it is possible to implement $\access$ in $O(1)$.
(A prior method than that using $\select$ is described by~\citet*{Vo:1998:CIF:290941.291011},
who adopted a byte-wise processing algorithm to accelerate skipping thorough the $H$ bitvector.)

We basically have to ``re-link'' together the high and low bits of an integer,
previously split up during the encoding phase.
The low bits $\ell_i$ are trivial to retrieve as we need to read the range of bits
$\ell_i = L[(i-1)\ell + 1, i\ell]$.
The retrieval of the high bits is, instead, more
complicated.
Since we write in unary how many integers share the same high part,
we have a bit set for every integer in $\seq$ and a zero for every distinct high part.
Therefore, to retrieve the high bits of the $i$-th integer,
we need to know how many zeros are present in the first $\select_{\bit{1}}(i)$ bits of $H$.
This quantity is evaluated on $H$ in $O(1)$ as $\select_{\bit{1}}(i)-i$.
Linking the high and low bits is as simple as:
$\access(i) = ((\select_{\bit{1}}(i)-i) \ll \ell) \mid \ell_i$,
where $\ll$ indicates the left shift operator and $\mid$ is the bitwise OR.

For example, to recover $\seq[4] = 13$, we first evaluate $\select_{\bit{1}}(4) - 4 = 5 - 4 = 1$
and conclude that the high part of $\seq[4]$ is the binary representation of 1, that is \bit{001}.
Finally, we access the low bits $L[10..12] = \bit{101}$ and re-link the two parts,
hence obtaining \bit{001.101}.

\parag{Supporting Successor queries}
The query $\suc(x)$, returning the smallest
integer $y$ of $\seq$ such that $y \geq x$, is supported in $O(1+\log(U/n))$ time as follows.
Let $h_x$ be the high bits of $x$. Then for $h_x > 0$, $i = \select_{\bit{0}}(h_x) - h_x + 1$ indicates that there are $i$ integers in $\seq$ whose high bits are less than $h_x$. On the other hand, $j = \select_{\bit{0}}(h_x + 1) - h_x$ gives us the position at which the elements having high bits greater than $h_x$ start. The corner case $h_x = 0$ is handled by setting $i = 0$.
These two preliminary operations take $O(1)$.
Now we can conclude the search in the range $\seq[i,j]$,
having \emph{skipped} a potentially large range of elements that, otherwise,
would have required to be compared with $x$.
We therefore determine the successor of $x$ by binary searching in this range which contains up to $U/n$ integers.
The time bound follows.


As an example, consider the query $\suc(30)$
over the example sequence from Table~\ref{tab:elias_fano}.
Since $h_{30} = 3$, we have $i = \select_{\bit{0}}(3) - 3 + 1 = 8$ and $j = \select_{\bit{0}}(4) - 3 = 9$. Therefore we conclude our search in the range $\seq[8,9]$ by returning $\suc(30) = \seq[9] = 36$.

In the specific context of inverted indexes, the
query {\suc} is called {\Nextgeq} (Next Greater-than or Equal-to)
and we are going to adopt this terminology in Section~\ref{sec:experiments}.
It should also be observed that~\citet*{moffat1996self} were the first
to explore the use of \emph{skip pointers} -- meta data aimed at accelerating the skipping
through blocks of compressed integers -- for faster query evaluation.

\parag{Partitioning the integers by cardinality}
One of the most pertinent characteristics of the Elias-Fano space bound
in Formula~\ref{ef_space} is that it only depends on two parameters,
i.e., the size $n$ of the sequence $\seq$ and the universe $U > \seq[n]$.
As already explained, inverted lists often present clusters of very similar
integers and Elias-Fano fails to exploit them for better compression
because it always uses a number of bits per integer at most equal to
$\lceil \log_2(U/n) \rceil + 2$, thus proportional to the logarithm
of the \emph{average gap} $U/n$ between the integers and \emph{regardless}
any skewed distribution.
(Note that also Golomb and Rice are insensitive to any deviation away from a
random selection of the integers.)
In order to better adapt to the distribution of the gaps between the integers,
we can partition the sequence, obtaining
the so-called \emph{partitioned Elias-Fano} ({\opt}) representation~\cite{2014:ottaviano.venturini}.

The sequence is split into $k$ blocks of variable length.
The first level of representation stores two sequences compressed with plain Elias-Fano:
(1) the sequence made up of the last elements $\{U_1,\ldots,U_k\}$ of the blocks,
the so-called \emph{upper-bounds} and
(2) the prefix-summed sequence of the sizes of the blocks.
The second level is formed, instead, by the representation of the blocks themselves, that can be again encoded with Elias-Fano.
The main advantage of this two-level representation, is that now the integers in the $i$-th block are encoded with a smaller universe, i.e., $U_i - U_{i-1}$, $i > 0$, thus improving the space with respect to the original Elias-Fano representation.
More precisely, each block in the second level is encoded with one among \emph{three} different strategies.
As already stated, one of them is Elias-Fano. The other two additional strategies come into play to overcome the space inefficiencies of Elias-Fano when representing \emph{dense} blocks.

%
%
%

Let consider a block and call $b$ its size, $M$ its universe respectively.
~\citet*{2013:vigna} observed that as $b$ approaches $M$ the space bound $b\lceil \log_2(M/b)\rceil + 2b$ bits becomes close to $2M$ bits. In other words, the closer $b$ is to $M$, the denser the block.
However, we can always represent the block with $M$ bits by writing the \emph{characteristic vector} of the block, that is a bitvector where the $i$-th bit is set if the integer $i$ belongs to the block.
Therefore, besides Elias-Fano, two additional encodings can be chosen to encode the block, according on the relation between $m$ and $b$.
The first one addresses the extreme case in which the block covers the whole universe, i.e., when $b = M$: in such case, the first level of the representation (upper-bound and size of the block) trivially suffices to recover each element of the block that, therefore, does not need to be represented at all.
The second case is used whenever the number of bits used by the Elias-Fano representation of the block is larger than $M$ bits: by doing the math, it is not difficult to see that this happens whenever $b > M/4$.
In this case we can encode the block with its characteristic bitvector using $M$ bits.
The choice of the proper encoding for a block is rather fundamental
for the practical space effectiveness of {\opt}.

Let us consider a simple example with $M = 40$ bits.
Suppose that the block is sparse, e.g., with $b = 5$.
Then, Elias-Fano takes $\lceil\log_2(40/5)\rceil + 2 = 5$ bits per element, whereas a characteristic vector representation would take $40/5 = 8$ bits per element.
In a dense case with, say, $b = 30$, a bitmap just takes $40/30 = 1.33$
bits per element, whereas Elias-Fano would take 3 bits per element.

Splitting the sequence into equally-sized block is clearly sub-optimal, since we cannot expect clusters of similar integers to be aligned with uniform partitions.
For such reason, an algorithm based on dynamic programming is presented by~\citet*{2014:ottaviano.venturini} that yields a partition whose cost in bits is at most $(1+\epsilon)$ times away from the optimal one taking $O(n \log \frac{1}{\epsilon})$ time and $O(n)$ space for any $0 < \epsilon < 1$.
Notice that the time complexity becomes $\Theta(n)$ when $\epsilon$ is constant.
In fact, the problem of determining the partition of minimum encoding cost can be seen as the problem of finding the path of minimum cost (shortest) in a complete, weighted and directed acyclic graph (DAG). This DAG has $n$ vertices, one for each integer of $\seq$, and $\Theta(n^2)$ edges where the cost $w(i,j)$ of edge $(i,j)$ represents the number of bits needed to represent $\seq[i,j]$. Each edge cost $w(i,j)$ is computed in $O(1)$ by just knowing the universe and size of the chunk $\seq[i,j]$.
By pruning the DAG it is possible to attain to the mentioned complexity by preserving the approximation guarantees~\cite{2014:ottaviano.venturini}.

\parag{Partitioning the integers by universe}
As already mentioned, we can opt for an arbitrary split between
the high and the low part of the Elias-Fano representation.
Partitioning the universe $U$ into chunks containing at most $2^{\ell}$
integers each, with $\ell = \lceil \log_2(U/n) \rceil$,
minimizes the space of the encoding~\cite{Elias74} but
a \emph{non-parametric} split -- independent from the
values of $U$ and $n$ -- is also possible.
Let us assume that $U \leq 2^{32}$ in the following.

For example, \emph{Roaring}~\cite{chambi2016better,lemire2016consistently,lemire2018roaring}
partitions $U$ into chunks spanning $2^{16}$ values each
and represents all the integers of the sequence falling into a chunk
in two different ways according to the cardinality of the chunk:
if the chunk contains less than 4096 elements, then it is considered
to be \emph{sparse} and represented as a sorted array of 16-bit integers;
otherwise it is considered \emph{dense} and encoded
as a bitmap of $2^{16}$ bits.
Lastly, very dense chunks can also be encoded with \emph{runs} if
advantageous.
A run is represented as a pair $(v,\ell)$ meaning that all the
integers $v \leq x \leq v + \ell$ belong to the chunk.

Inspired by the van Emde Boas tree~\cite{Boas75,Boas77},
the \emph{Slicing}~\cite{slicing} data structure recursively
slices the universe of representation 
in order to better adapt to the
distribution of the integers being compressed.
Differently from Roaring, a \emph{sparse} sparse chunk is
further partitioned into at most $2^8$ blocks of $2^8$ elements each.
Therefore, a non-empty universe slice of $2^{16}$ elements can be either:
represented with a bitmap of $2^{16}$ bits (dense case);
represented implicitly if the slice contains
all the possible $2^{16}$ elements (full case);
or it is recursively partitioned into smaller slices of
$2^8$ elements each.
Finally, each non-empty slice of $2^8$ elements is
encoded with a sorted array of 8-bit integers (sparse case);
or with a bitmap of $2^{8}$ bits (dense case).
The idea of a hybrid compression scheme with hierarchical bit-vectors and sorted-arrays
(that can be further compressed) was first proposed by~\citet{fraenkel1986improved}.

%
%

It should be noted that all the partitioning strategies we have
described in this section, namely partitioned Elias-Fano ({\pef}),
Roaring and Slicing, exploit the same idea to attain to good space
effectiveness: look for dense regions to be encoded with bitmaps
and use a different mechanism for sparse regions.
While {\pef} achieves this goal by splitting the sequence
\emph{optimally} by cardinality, Roaring and Slicing partition
the universe of representation \emph{greedily}, hence maintaining
the property that all partitions are represented using the \emph{same}
universe.
As we will better see in Section~\ref{sec:experiments}, these different
partitioning paradigms achieve different space/time trade-offs.

\begin{figure}
\centering
\includegraphics[scale = 0.74]{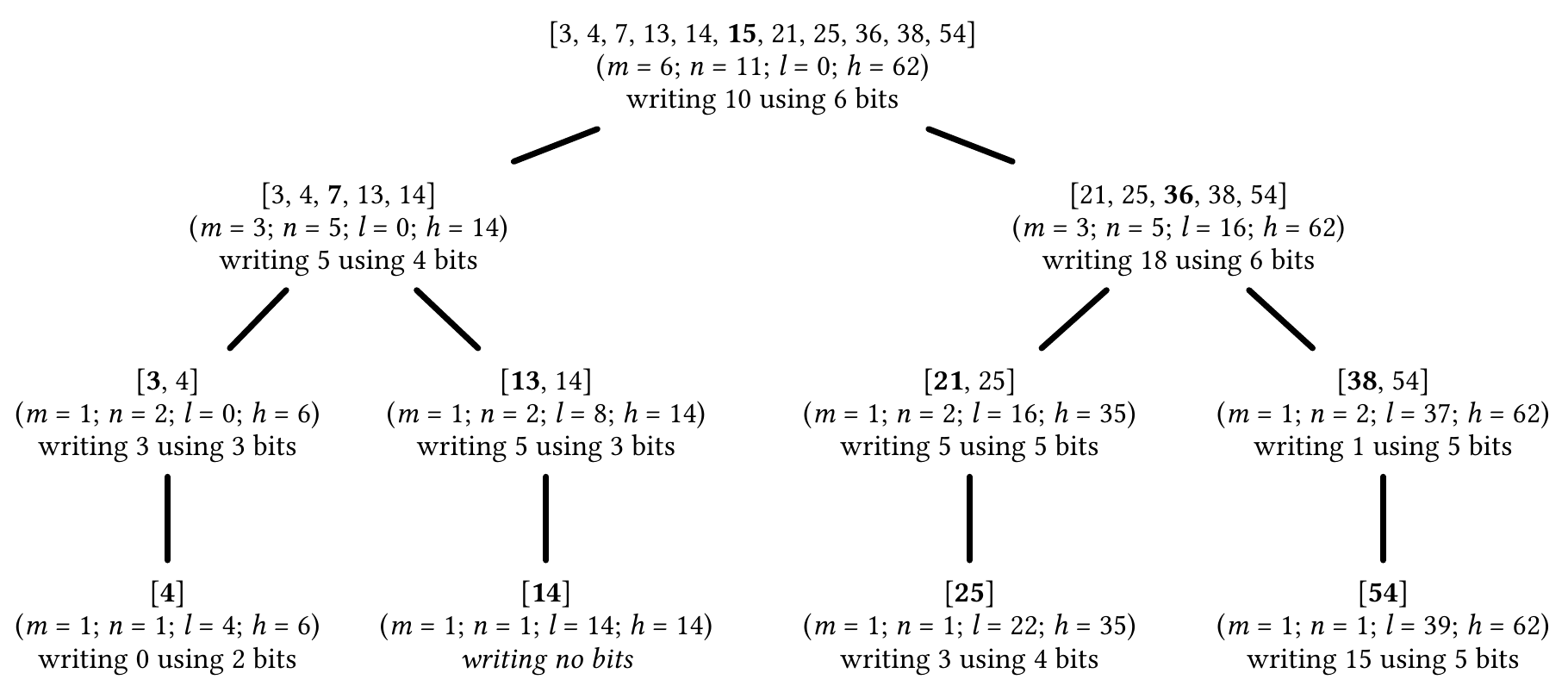}
\caption{The recursive calls performed by the Binary Interpolative Coding
algorithm when applied to the sequence
$[$3, 4, 7, 13, 14, 15, 21, 25, 36, 38, 54$]$
with initial knowledge of lower and upper bound values $l = 0$ and $h = 62$.
In bold font we highlight the middle element being encoded.
}
\label{fig-bic}
\end{figure}

\subsection{Interpolative}\label{subsec:BIC}

The \emph{Binary Interpolative Code} ({\bic})
invented by~\citet*{1996:moffat.stuiver,2000:moffat.stuiver}
represents a sorted integer sequence without requiring the computation of its gaps.
The key idea of the algorithm is to exploit the order
of the already-encoded elements to compute the number of bits needed
to represent the elements that will be encoded next.

At the beginning of the encoding phase, suppose we are specified
two quantities $l \leq \seq[1]$ and $h \geq \seq[n]$.
Given such quantities, we can encode the element
in the middle of the sequence, i.e., $\seq[m]$ with $m = \lceil n / 2 \rceil$,
in some appropriate manner, knowing
that $l \leq \seq[m] \leq h$.
For example, we can write $\seq[m] - l - m + 1$ using just
$\lceil \log_2(h - l - n + 1) \rceil$ bits.
After that, we can apply the same step to both halves $\seq[1, m-1]$
and $\seq[m+1, n]$ with \emph{updated knowledge} of lower and upper values
$(l,h)$ that are set to $(l, \seq[m] - 1)$ and $(\seq[m] + 1, h)$
for the left and right half respectively.
Note that whenever the condition $l+n-1=h$ is satisfied, a ``run'' of
consecutive integers is detected: therefore, we can stop recursion and emit
no bits at all during the encoding phase.
When the condition is met again during the decoding phase,
we simply output the values $l,l+1,l+2,\ldots,h$.
This means that {\bic} can actually use codewords of 0 bits
to represent more than one integer, hence attaining to a rate
of less than one bit per integer -- a remarkable property that makes
the code very succinct for highly clustered inverted lists.

We now consider an encoding example applied to the sequence
$[$3, 4, 7, 13, 14, 15, 21, 25, 36, 38, 54, 62$]$.
As it is always safe to choose $l = 0$ and $h = \seq[n]$,
we do so, thus at the beginning of the encoding phase we
have $l = 0$ and $h = 62$. Since we set $h = \seq[n]$,
the last value of the sequence is first encoded and we
process $\seq[1..n-1]$ only.
Fig.~\ref{fig-bic} shows the sequence of recursive calls
performed by the encoding algorithm oriented as a binary tree.
At each node of the tree we report
the values assumed by the quantities
$m$, $n$, $l$ and $h$, plus the processed subsequence
and the number of bits needed to encode the middle element.
By \emph{pre-order} visiting the tree, we obtain the sequence
of written values, that is $[$10, 5, 3, 0, 5, 18, 5, 3, 1, 15$]$
with associated codeword lengths
$[$6, 4, 3, 2, 3, 6, 5, 4, 5, 5$]$.
Note that the value in the second leaf of the tree, i.e., $\seq[5] = 14$,
is encoded with 0 bits given that both $l$ and $h$ are equal to 14.



However, the encoding process obtained by the use of simple
binary codes as illustrated in Fig.~\ref{fig-bic}
is wasteful.
In fact, as discussed in the original work~\cite{1996:moffat.stuiver,2000:moffat.stuiver},
more succinct encodings can be achieved with a
minimal binary encoding (recall the definition at the end of Section~\ref{sec:unary_binary}).
More precisely, when the range $r > 0$ is specified, all values
$0 \leq x \leq r$ are assigned fixed-length codewords of size
$\lceil \log_2(r+1) \rceil$ bits. But the more $r$ is distant
from $2^{\lceil \log_2(r+1) \rceil}$ the more this allocation
of codewords is wasteful because
$c = 2^{\lceil \log_2(r+1) \rceil} - r - 1$
codewords can be made 1 bit shorter without loss of unique
decodability.
Therefore we proceed as follows.
We identify the range of smaller codewords, delimited by
the values $r_{l}$ and $r_{h}$, such that every value
$x \leq r$ such that
$r_{l} < x < r_{h}$ is assigned a shorter
$(\lceil \log_2(r+1) \rceil - 1)$-bit codeword
and every value outside this range is assigned
a longer one of $\lceil \log_2(r+1) \rceil$ bits.
To maintain unique decodability, we first always read
$\lceil \log_2(r+1) \rceil - 1$ bits and interpret these as the
value $x$.
Then we check if condition $r_{l} < x < r_{h}$ is satisfied:
if so, we are done; otherwise, the codeword must be
extended by 1 bit.
In fact, in a \emph{left-most} minimal binary code assignment,
the first $c$ values are assigned the shorter codewords,
thus $r_{h} = 2^{\lceil \log_2(r+1) \rceil} - r - 1$
(and we only check whether $x < r_{h}$).
In a \emph{centered} minimal binary code assignment,
the values in the centre of the range are assigned the
shorter codewords, thus
$(r_{l}, r_{h}) = (\lfloor r / 2 \rfloor - \lfloor c / 2 \rfloor - 1,
\lfloor r / 2 \rfloor + \lfloor c / 2 \rfloor + 1)$
if $r$ is even, or
$(r_{l}, r_{h}) = (\lfloor r / 2 \rfloor - \lfloor c / 2 \rfloor,
\lfloor r / 2 \rfloor + \lfloor c / 2 \rfloor + 1)$
if $r$ is odd.
The rationale behind using centered minimal codes
is that a reasonable guess is to assume the middle
element to be about half of the upper bound.
As already noted, we remark that the specific assignment of codewords
is irrelevant and many assignments are possible:
what matters is to assign correct lengths
and maintain the property of unique decodability.


It is also worth mentioning that the \emph{Tournament code} developed by~\citet*{teuhola2008tournament}
is very related to {\bic}.


%

\subsection{Directly-addressable codes}

\citet*{brisaboa2013dacs} introduced a representation for a list of integers
that supports random access to individual integers -- called \emph{directly-addressable code} (\textsf{DAC}) --
noting that this is not generally possible for many of the representations
described in Section~\ref{sec:integer} and~\ref{sec:list}.
They reduced the problem of random access to the one of \emph{ranking} over a bitmap.
Given a bitmap $B[1..n]$ of $n$ bits, the query $\rank_b(B,i)$ returns the number of $b$ bits in
$B[1,i]$, for $i \leq n$.
For example, if $B = \bit{010001101110}$ then $\rank_{\bit{1}}(6) = 2$ and $\rank_{\bit{0}}(8) = 5$.
Rank queries can be supported in $O(1)$ by requiring only $o(n)$ additional bits~\cite{JacobsonPhD89,ClarkPhD96,GGMN05}.

Each integer in the list is partitioned into $(b+1)$-bit chunks. Similarly to Variable-Byte,
$b$ bits are dedicated to the representation of the integer and the control bit indicates whether
another chunk follows or not.
All the first $n$ $b$-bit chunks of every integer are grouped together in a codeword stream $C_1$
and the control bits form a bitmap $B_1[1..n]$ of $n$ bits.
If the $i$-th bit is set in such bitmap, then the $i$-th integer in the sequence
needs a second chunk, otherwise a single chunk is sufficient.
Proceeding recursively, all the second $m \leq n$ chunks are concatenated together in $C_2$ and
the control bits in a bitmap $B_2[1..m]$ of $m$ bits. Again, the $i$-th bit of such bitmap is set if the
$i$-th integer with at least two chunks needs a third chunk of representation.
In general, if $U$ is the maximum integer in the list, there are at most $\lceil \log_2(U + 1) / b \rceil$
levels (i.e., streams of chunks).

As an example,
the sequence $[$2, 7, 12, 5, 13, 142, 61, 129$]$ is encoded with $b=3$
as follows:
$B_1 = \bit{0.0.1.0.1.1.1.1}$;
$B_2 = \bit{0.0.1.0.1}$;
$B_3 = \bit{0.0}$;
$C_1 = \bit{010.111.100.101.101.110.011.001}$;
$C_2 = \bit{001.001.001.110.000}$;
$C_3 = \bit{010.010}$.

Accessing the integer $x$ in position $i$ reduces to a sequence of $c-1$ $\rank_1$ operations over
the levels' bitmaps, where $c \geq 1$ is the number of $(b+1)$-bit chunks of $x$, that is
$\lceil \log_2(x + 1) / b \rceil$.
Now, for $k=1..c$, we repeat the following step:
(1) retrieve the $i$-th chunk from the $C_k$ in constant time given that all chunks are $b$ bits long;
(2) if $B_k[i] = 0$, we are done; otherwise $j = \rank_1(B_k,i)$ gives us the number of integers
(in the level $k+1$) that have more than $k$ chunks, so we set $i = j$ and repeat.
For example, $\access(5)$ is resolved as follows on our
example sequence. We retrieve $C_1[5] = \bit{101}$; since $B_1[5]=\bit{1}$,
we compute $\rank_1(B_1,5)=2$. Now we retrieve $C_2[2]=\bit{001}$ and given $B_2[2]=\bit{0}$, we stop
by returning the integer $C_2[2]\bit{.}C_1[5] = \bit{001}.\bit{101}$, that is 13.

Lastly, nothing prevents from changing the value of $b$ at each level of the data structure.
For this purpose, the authors of DAC present an algorithm, based on dynamic programming,
that finds such optimal values for a given list.

\subsection{Hybrid approaches}\label{subsec:optvb}

Hybrid approaches are possible by using different compressors
to represent the blocks of a list.
For example, given a query log, we can collect access statistics
at a block-level granularity,
namely how many times a block is accessed during query processing,
and represent rarely-accessed blocks with more space-efficient compressor;
vice versa frequently-accessed blocks are encoded with more
time-efficient compressor~\cite{2015:ottaviano.tonellotto.ea}.
This hybrid strategy produces good space/time trade-offs.

~\citet*{TKDE19} show that a list of $n$ sorted integers can be
optimally partitioned into variable-length blocks whenever
the chosen representation for each block is given by either:
(1) any compressor described in Section~\ref{sec:integer},
namely a \emph{point-wise} encoder, or
(2) the characteristic vector of the block.
From Section~\ref{EF} we recall that, given a block of universe $m$,
the characteristic vector representation of the block is given by
a bitmap of $m$ bits where the $i$-th bit is set if the integer
$i$ belongs to the block.
By exploiting the fact that the chosen encoder is point-wise, i.e.,
the number of bits needed to represent an integer solely depends on the
integer itself rather than the block where it belongs to,
it is possible to devise an algorithm that finds an \emph{optimal} partitioning
in $\Theta(n)$ time and $O(1)$ space.
The constant factor hidden by the asymptotic notation is very small,
making the algorithm very fast in practice.

\subsection{Entropy coding: Huffman, Arithmetic, and Asymmetric Numeral Systems}\label{sec:entropy_coding}

In this section we quickly survey the most famous
\emph{entropy coding} techniques -- Huffman~\cite{Huffman52},
Arithmetic coding~\cite{rissanen1976generalized,pasco1976source,rissanen1979arithmetic,Moffat:1998:ACR:290159.290162},
and Asymmetric Numeral Systems ({\ans})~\cite{duda09,duda13,duda15}.
Although some authors explored the use of these techniques for index compression,
especially Huffman~\cite{Jakobsson78,bookstein1989construction,fraenkel1985novel,moffat1992parameterised}
and {\ans}~\cite{ANS1,ANS2} (see Section~\ref{subsec:ANS-based}),
they are usually not competitive in terms of efficiency
and implementation simplicity
against the compressors we have illustrated in the previous sections,
making them a hard choice for practitioners.
An in-depth treatment of such techniques is, therefore, outside the scope of this article
and the interested reader can follow the references to the individual papers we include here.
The survey by~\citet{huffman_coding} about Huffman coding
also contains descriptions of Arithmetic Coding and {\ans} (Section 5.1 and 5.2 of that article, respectively).

We first recall the definition
of \emph{entropy}, a tool introduced by~\citet*{Shannon48}.
He was concerned with the problem of defining the \emph{information content} of a discrete random variable
$\mathcal{X} : \Sigma \rightarrow \mathbb{R}$, with distribution $\pr(s) = \pr\{\mathcal{X} = s\}$, $s \in \Sigma$.
He defined the entropy of $\mathcal{X}$ as
$H = \sum_{s \in \Sigma}[\pr(s)\log_2(1/\pr(s))]$ bits.
The quantity $\log_2(1/\pr(s))$ bits is also called the \emph{self-information} of the symbol $s$
and $H$ represents the average number of bits we need to encode each value of $\Sigma$.
Let now $\seq$ be a sequence of $n$ symbols drawn from an alphabet $\Sigma$.
(In the context of this article, the symbols will be integer numbers.)
Let also $n_s$ denote the number of times the symbol $s$ occurs in $\seq$.
Assuming empirical frequencies as probabilities~\cite{Papoulis91} (the larger is $n$, the better the approximation), i.e., $\pr(s) \approx n_s / n$, we can consider $\seq$ as a random variable assuming value $s$ with probability $\pr(s)$.
In this setting, the entropy of the sequence $\seq$ is
$H_0 = 1/n\sum_{s \in \Sigma}[n_s\log_2(n/n_s)]$ bits,
also known as the $0$-th order (or \emph{empirical}) entropy of $\seq$.
In particular, the quantity $nH_0$ gives a theoretic lower bound on the average number of bits
we need to represent $\seq$ and, hence, to the output size of \emph{any} compressor
that encodes each symbol of $\seq$ with a fixed-length codeword.

\begin{figure}[t]
\begin{minipage}[t]{.25\textwidth}
\vspace{0pt}
\scalebox{\mytablescale}{\hspace{-3cm}\begin{tabular}{cccl}
\toprule

symbols & weights & lengths & codewords \\

\midrule

2 & 8 & 2 & \bit{00} \\
5 & 7 & 2 & \bit{01} \\
6 & 2 & 3 & \bit{100} \\
7 & 2 & 3 & \bit{101} \\
1 & 2 & 4 & \bit{1100} \\
3 & 2 & 4 & \bit{1101} \\
4 & 1 & 4 & \bit{1110} \\
8 & 1 & 4 & \bit{1111} \\

\bottomrule
\end{tabular}
}
\end{minipage}
\hspace{0.5cm}
\begin{minipage}[t]{.25\textwidth}
\vspace{0pt}
\includegraphics[scale=0.5]{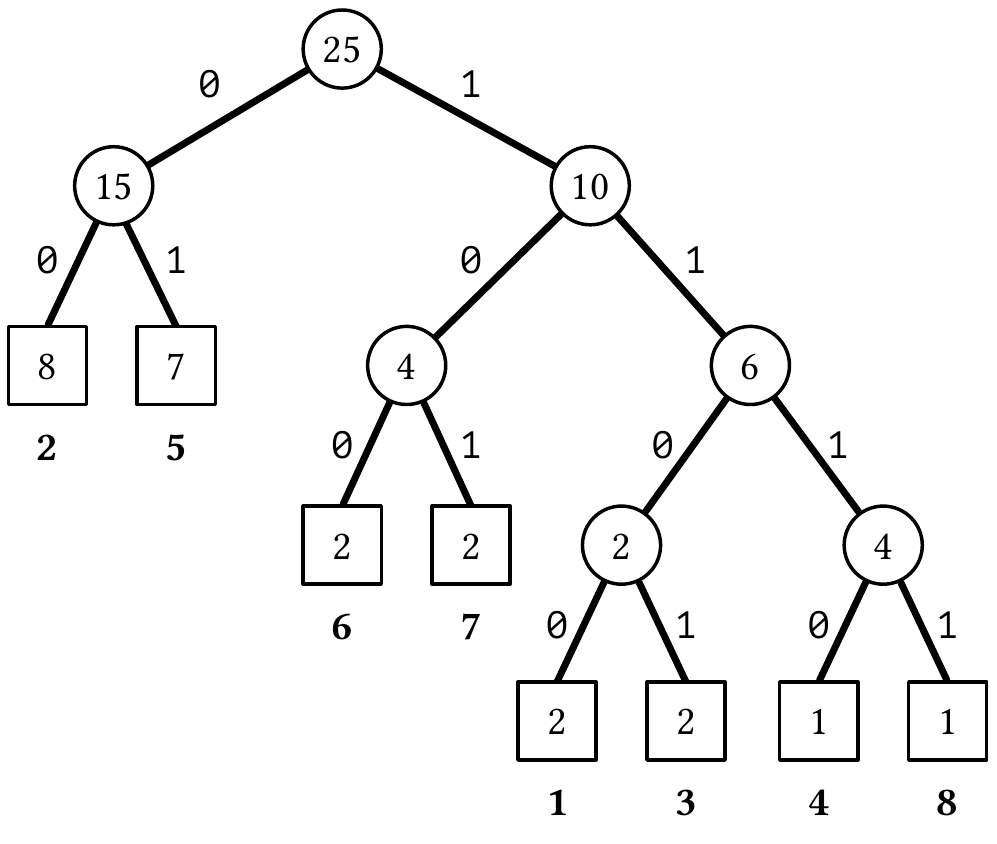}
\end{minipage}
\caption{An example of Huffman coding applied to a sequence of size 25 with symbols 1..8
and associated weights $[2,8,2,1,7,2,2,1]$.
\label{fig:huffman}}

\end{figure}

\parag{Huffman}
It is standard to describe the Huffman's algorithm in terms of a binary tree.
In this logical binary tree,
a leaf corresponds to a symbol to be encoded with associated symbol occurrence -- its
weight -- and
an internal node stores the sum of the weights of its children.
The algorithm maintains a \emph{candidate set} of tree nodes from which, at each step:
(1) the two nodes with smallest weight are selected;
(2) they are merged together into a new parent node whose weight
is the sum of the weights of the two children;
(3) the parent node is added to the candidate set.
The algorithm repeats this merging step until only the root of the tree (whose weight
is the length of the sequence $\seq$) is left in the candidate set.
Fig.~\ref{fig:huffman} shows an example of Huffman coding.
It is important to mention that, in practice, the decoding process
does \emph{not} traverse any tree.
An elegant variation of the algorithm -- known as \emph{canonical Huffman} --
allows fast decoding by using lookup tables as we similarly illustrated in Section~\ref{sec:encoding_decoding}.
Again, ~\citet*{huffman_coding} provides all details.

Now, let
$L$ be the \emph{average} Huffman codeword length.
Two of the most important properties of Huffman coding are:
(1) $L$ is \emph{minimum} among all possible prefix-free codes;
(2) $L$ satisfies $H_0 \leq L < H_0 + 1$.
The first property means that Huffman coding produces an \emph{optimal} code
for a given distribution of the integers.
(The precursor of the Huffman's algorithm is the less-known \emph{Shannon-Fano}
algorithm that was
independently proposed by~\citet*{Shannon48} and~\citet*{Fano49}, which, however,
does not always produce an optimal code.)
The second property suggests that
an Huffman code can loose up to 1 bit compared to the entropy $H_0$
because it
requires \emph{at least} 1 bit
to encode a symbol (as any other prefix-free code),
thus if $H_0$ is large, the extra bit lost is negligible
in practice; otherwise the distribution of probabilities is skewed and Huffman
looses a significant space compared to the entropy of the source.

%
%


Now, it should be clear why Huffman may \emph{not} be an appropriate
choice for inverted index compression.
Applying Huffman to the compression of the integers in inverted lists
means that its alphabet of representation is too large, thus making
the mere description of the code outweigh the cost of representing
very sparse inverted lists.
The same reason applies if we try to use the code
to compress the gaps
between successive integers: the largest gap could be as large as
the largest integer in the sequence.


\parag{Arithmetic}
The first concept of Arithmetic coding was introduced by Elias before
1963 according to Note 1 on page 61 in the book by~\citet*{abramson1963information}.
However, the method requires infinite precision arithmetic and,
because of this, it remained unpublished.
The first practical implementations
were designed during 1976 by~\citet*{rissanen1976generalized} and~\citet*{pasco1976source},
and later refined by~\citet*{rissanen1979arithmetic}.
A more recent efficient implementation is described by~\citet{Moffat:1998:ACR:290159.290162}.
The method offers higher compression ratios
than Huffman's,
especially on highly skewed distributions,
because it is \emph{not} a prefix-free code, so it does not require at least
one bit to encode a symbol. Indeed a single bit may correspond to more
than one input symbol.
However, Huffman codes are faster to decode; Arithmetic does not permit
to decode an output stream starting from an arbitrary position, but only
sequential decoding is possible.

Given a sequence of symbols $\seq = [s_1..s_n]$, the main idea behind the method works as follows.
The interval $[0,1)$ is partitioned into $|\Sigma|$ segments of length proportional to the probabilities of the symbols.
Then the subinterval corresponding to $s_1$, say $[\ell_1, r_1)$, is chosen and the same
partitioning step is applied to it.
The process stops when all input symbols have been processed
and outputs a single real number $x$ in $[\ell_n, r_n)$,
that is the interval associated to the last input symbol $s_n$.
Then the pair $(x, n)$ suffices to decode the original input sequence $\seq$.


It can be shown that Arithmetic coding takes at most $nH_0 + 2$ bits to encode
a sequence $\seq$ of length $n$. This means that the overhead with respect to
the empirical entropy $H_0$ is only of $2/n$ bits per symbol, thus negligible
for basically all practical values of $n$.
As already pointed out, Arithmetic coding requires infinite precision that can
be very costly to be approximated. In fact, a practical implementation~\cite{witten1987arithmetic} using
approximated arithmetic can take up to $nH_0 + \frac{2}{100}n$ bits, thus
having $0.02$ bits of loss per symbol rather than $2/n$.

\parag{Asymmetric Numeral Systems}
Asymmetric Numeral Systems ({\ans}) is a family of entropy coding algorithms,
originally developed by~\citet*{duda09,duda13}, which approaches the compression ratio
of Arithmetic coding with a decompression speed comparable with the one of Huffman~\cite{duda15}.
The basic idea of {\ans} is to represent a {sequence}
of symbols with a natural number $x$.

Let us consider a concrete example~\cite{ANS2} with an alphabet of 3 symbols only,
namely $\{a,b,c\}$ and assuming that $\pr(a)=1/2$, $\pr(b)=1/3$ and $\pr(c)=1/6$.
In order to derive the encoding of a sequence of symbols, a \emph{frame} $f[1..m]$
of symbols is constructed, having $1/2$ of the entries equal to $a$, $1/3$ equal to $b$ and $1/6$
equal to $c$.
For example, one such frame could be $f[1..6] = [aaabbc]$, but other symbol permutations
with possibly larger $m$ are possible as well.
The frame determines a table that is used to map
a sequence of symbols to an entry in the table.
Refer to Table~\ref{fig:ans1} for an example with the frame $f[1..6] = [aaabbc]$.
The entries in the table are the natural numbers assigned incrementally in the order determined
by the frame. For example, since the first three symbols in the frame are $aaa$, the first numbers
assigned to $a$'s row are 1, 2 and 3.
The next two symbols are $bb$, so $b$'s row gets 4 and 5.
The last symbol in the frame is $c$, so the first entry in $c$'s row is 6.
The process now proceed in cycles, thus placing 7, 8 and 9 in $a$'s row; 10 and 11 in $b$'s row,
a final 12 in $c$'s row, and so on (first 10 columns are shown in the table).
Table~\ref{fig:ans2} shows an example for another distribution of the symbols,
constructed using a frame $f[1..4] = [caba]$.

\begin{table}
\centering
\caption{Two example of {\ans} encoding table, with respectively frame $f[1..6] = [aaabbc]$ and $f[1..4] = [caba]$.}
\subfloat[]{
	\scalebox{0.8}{
    \begin{tabular}{c c c c c c c c c c c c}
\toprule

$\Sigma$ & $\pr$ & \multicolumn{10}{c}{codes}  \\

\midrule

$a$ & $1/2$ & 1 & 2 & 3 & 7 & 8 & 9 & 13 & 14 & 15 & 19 \\
$b$ & $1/3$ & 4 & 5 & 10 & 11 & 16 & 17 & 22 & 23 & 28 & 29 \\
$c$ & $1/6$ & 6 & 12 & 18 & 24 & 30 & 36 & 42 & 48 & 54 & 60 \\

\bottomrule

& & 0 & 1 & 2 & 3 & 4 & 5 & 6 & 7 & 8 & 9
\end{tabular}
}
    \label{fig:ans1}
}
\hspace{0.3cm}
\subfloat[]{
	\scalebox{0.8}{
    \begin{tabular}{c c c c c c c c c c c c}
\toprule

$\Sigma$ & $\pr$ & \multicolumn{10}{c}{codes}  \\

\midrule

$a$ & $1/2$ & 2 & 4 & 6 & 8 & 10 & 12 & 14 & 16 & 18 & 20 \\
$b$ & $1/4$ & 3 & 7 & 11 & 15 & 19 & 23 & 27 & 31 & 35 & 39 \\
$c$ & $1/4$ & 1 & 5 & 9 & 13 & 17 & 21 & 25 & 29 & 33 & 37 \\

\bottomrule

& & 0 & 1 & 2 & 3 & 4 & 5 & 6 & 7 & 8 & 9
\end{tabular}
}
    \label{fig:ans2}
}
\end{table}

Now, consider the sequence $caa$ and let us determine its {\ans} code with the table
in Fig.~\ref{fig:ans1}. We make use of the \emph{transition function} defined by the table $T$
itself as $\var{state}^{\prime} = T[s, \var{state}]$ which, given a symbol $s$ and a \var{state} value,
produces the next $\var{state}^{\prime}$ of the encoder.
At the beginning we set $\var{state} = 0$, thus for the given sequence $caa$ the \var{state} variable
assumes values $0 \rightarrow 6 \rightarrow 13 \rightarrow 26$ (last value not shown in the table).
The code assigned to the sequence is, therefore, the integer 26.
For the sequence $acb$ under the encoding table in Table~\ref{fig:ans2} we generate, instead,
the transitions $0 \rightarrow 2 \rightarrow 9 \rightarrow 39$, thus the assigned code is 39.
Decoding reverts this process. For example, given 39 we know that the last symbol of the encoded
sequence must have been $b$ because 39 is found on the second row of the table. The value
is in column 9, which is found in column 2 in the third row that corresponds to the $c$ symbol.
Finally, the column number 2 is found in $a$'s row, thus we emit the message $acb$.

\section{Index Compressors}\label{sec:index}

This section is devoted to approaches that look for
regularities among \emph{all} the lists in the inverted index.
In fact, as already motivated at the beginning of Section~\ref{sec:list},
the inverted index naturally presents some amount of redundancy
in that many sub-sequences of integers are shared between the lists.
Good compression
can be achieved by exploiting this correlation,
usually at the expense of a reduced
processing efficiency.

\subsection{Clustered}
\citet*{2017:pibiri.venturini} propose a clustered index representation.
The inverted lists are grouped into clusters of ``similar'' lists, i.e.,
the ones sharing as many integers as possible.
Then for each cluster, a \emph{reference list} is synthesized with respect to which
all lists in the cluster are encoded.
More specifically, the integers belonging to the intersection between the cluster reference list
and a list in the cluster are represented as the positions they occupy within the reference list.
This makes a big improvement for the cluster space, since each intersection can be re-written
in a much smaller universe.
Although \emph{any} compressor can be used to represent the
intersection and the residual segment of each list,
the authors adopt partitioned Elias-Fano;
by varying the size of the reference lists, different time/space trade-offs can be obtained.

\subsection{ANS-based}\label{subsec:ANS-based}

In Section~\ref{sec:entropy_coding} we have seen an example of the {\ans} method
developed by~\citet*{duda09,duda13}.
As we have already observed in that section,
the alphabet size may be too large for representing the integers in inverted indexes.
Even the largest gap may be equal to the
number of documents in the collection, which is usually several order to magnitudes
larger than, for example, the (extended) ASCII alphabet.
For this reason,~\citet*{ANS1} describe several adaptations of the base {\ans} mechanism
tailored for effective index compression.
In order to reduce the alphabet size, they perform a preprocessing step with Variable-Byte
to reduce the input list to a sequence of bytes and then apply {\ans} (\textsf{VByte+ANS}).
Local variability can be instead captured by using 16 different {\ans} models, each selected
using a 4-bit selector in the spirit of the Simple approach described in Section~\ref{subsec:simple}
(\textsf{Simple+ANS}).
Another variant is obtained by dividing a list into blocks and encoding each block with the
most suitable model, chosen among 16 possibilities according to a selected block statistic
e.g., its maximum value (\textsf{Packed+ANS}).


\subsection{Dictionary-based}\label{subsec:dint}

\begin{figure}
\centering
\includegraphics[scale=0.73]{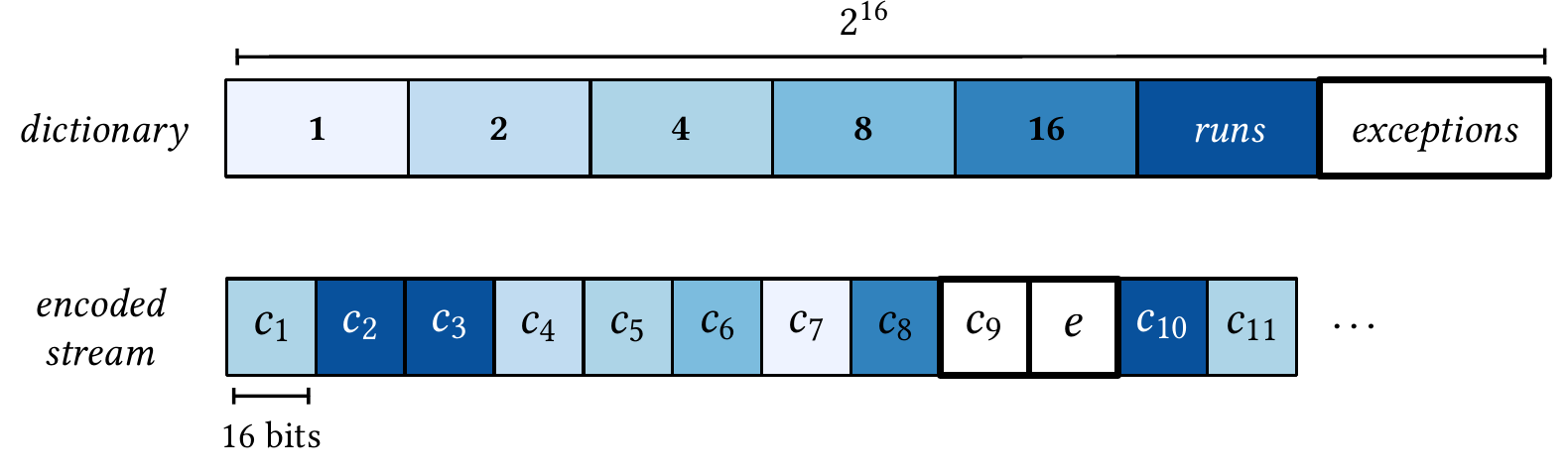}
\caption{A dictionary-based encoded stream example,
where dictionary entries corresponding
to $\{1,2,4,8,16\}$-long integer
patterns, runs and exceptions, are labelled with different shades.
Once provision has been made for such a dictionary structure,
a sequence of gaps can be modelled as a sequence of
codewords $\{c_k\}$, each being a reference to a dictionary entry,
as represented with the \emph{encoded stream} in the picture.
Note that, for example, codeword $c_9$ signals an exception,
therefore the next symbol $e$ is decoded using an escape mechanism.
\label{fig:dint}}
\end{figure}

\citet*{DINT} show that inverted indexes can be effectively compressed using
a dictionary-based approach. Their technique
-- named Dictionary of INTeger sequences ({\dint}) --
builds on the observation that
patterns of gaps are highly repetitive across the whole inverted index.
For example, the pattern $[1,1,2,1]$ of 4 gaps can be very repetitive.
Therefore, a dictionary storing the \emph{most frequent} $2^b$ patterns, for some $b > 0$, can be constructed.
Note that, in general,
the problem of building a dictionary that minimizes the
number of output bits when sequences symbols are coded as references
to its entries is NP-hard~\cite{ss82jacm}.
More specifically, an integer list can be modelled as a sequence of $b$-bit codewords,
each codeword corresponding to a dictionary pattern.
Fig.~\ref{fig:dint} illustrates the approach.
This representation has the twofold advantage of:
(1) requiring $b$ bits to
represent a pattern (thus, potentially, several integers);
(2) decoding of a pattern requires just a lookup in the dictionary.
In their investigation, patterns of size 1, 2, 4, 8 and 16 are considered, with
$b=16$ to avoid bit-level manipulations and allow very fast decoding.

A detail of crucial importance is to take advantage of the presence of
\emph{runs} of 1s, hence reserving some special entries in the dictionary to
encode runs of different sizes, such as 32, 64, 128, and 256.
A dictionary entry must also be reserved to signal the presence
of an \emph{exception} -- an integer not sufficiently frequent
to be included in the dictionary and represented via an escape mechanism
(e.g., Variable-Byte or a plain 32-bit integer).
Moreover, compacting the dictionary has the potential of letting the dictionary
fit in the processor cache, hence speeding up the decoding process
thanks to a reduced number of cache misses.
Lastly, once the dictionary is built, a shortest-path computation
suffices to find the \emph{optimal encoding} of a list for that specific
dictionary.


Other authors have instead advocated the use of \emph{Re-Pair}~\cite{1999:larsson.moffat}
to compress the gaps between the integers of
inverted lists~\cite{cfn09arxiv,2016:claude.farina.ea}.
Re-Pair uses a grammar-based approach to generate a dictionary
of sequences of symbols. Description of this algorithm is outside the
scope of this article.

\section{Further readings}\label{sec:further}

Besides the individual papers listed in the bibliography,
we mention here previous efforts in summarizing encoding
techniques for integers/integer sequences.
The book by~\citet*{mg} is the first, to the best of our knowledge,
that treats compression and indexing data as a unified problem, by
presenting techniques to solve it efficiently.
~\citet*{fenwick2003universal} and~\citet*{Salomon07} provide
a vast and deep coverage of variable-length codes.
The survey by~\citet*{ZobelM06} covers more than 40 years of
academic research in Information Retrieval and gives an
introduction to the field, with Section 8 dealing with efficient
index representations.
~\citet*{moffat2002compression},~\citet*{Moffat16},~\citet*{EBDT2018}
describe
several of the techniques illustrated in this article;
~\citet*{1999:williams.zobel}, ~\citet{scholer2002compression} and
~\citet*{trotman2003compressing} experimentally evaluate many of them.

Other approaches not described in this article include:
an adaptation of \emph{Front Coding}~\cite{mg}
for compressing text, seen as formed by quadruples holding
document, paragraph,
sentence, and word number~\cite{cfk88sigir};
the use of general-purpose compression libraries,
such as \textsf{ZStd}\footnote{\url{http://www.zstd.net}}
and \textsf{XZ}\footnote{\url{http://tukaani.org/xz}},
for encoding/decoding of inverted lists~\cite{petri2018compact}.


\section{Experiments}\label{sec:experiments}

In this section of the article, we report
on the space effectiveness and time efficiency
of different inverted index representations.
Specifically, space effectiveness is measured as the average
number of bits dedicated to the representation of
a document identifier; time efficiency is assessed in terms of the
time needed to perform
sequential decoding, intersection, and union of inverted lists.
For the latter two operations,
we focus on materializing the \emph{full} results set, without
any ranking or dynamic pruning mechanism~\cite{Wand,BlockMaxWand}
being applied.


We do not aim at being exhaustive here but
rather compare some selected representations and
point the interested reader to the code repository
at \url{https://github.com/jermp/2i_bench}
for further comparisons.


\parag{Tested index representations}
We compare the 12 different configurations,
summarized in Table~\ref{tab:strategies}.
We report some testing details of such configurations.
The tested Rice implementation (Section~\ref{sec:rice})
specifies the Rice parameter $k$ for each block of integers,
choosing the value of $k \in [1,4]$ giving the best space effectiveness.
Two bits per block suffices to encode the value of $k$.
Also, we write the quotient of the Rice
representation of an integer in $\gamma$ rather than in unary,
as we found this to give a better space/time trade-off than regular Rice.
Variable-Byte uses the SIMD-ized decoding algorithm
devised by~\citet{2015:plaisance.kurz.ea} and called Masked-VByte.
Interpolative ({\bic}) uses
\emph{leftmost minimal} binary codes.
The tested version of {\dint}
uses a single packed dictionary and optimal block parsing.
In Roaring, extremely dense chunks are represented with runs.

\begin{table}[t]
\caption{The different tested index representations.}
\scalebox{\mytablescale}{\begin{tabular}{l c c c l
}
\toprule

Method & Partitioned by & SIMD & Alignment & Description
\\

\midrule


{\vb} &
cardinality &
yes &
byte &
fixed-size partitions of 128
\\

{\optvb} &
cardinality &
yes &
bit &
variable-size partitions
\\

{\interp} &
cardinality &
no &
bit &
fixed-size partitions of 128
\\

{\cdelta} &
cardinality &
no &
bit &
fixed-size partitions of 128
\\


{\rice} &
cardinality &
no &
bit &
fixed-size partitions of 128
\\

{\opt} &
cardinality &
no &
bit &
variable-size partitions
\\

{\dint} &
cardinality &
no &
\,\,\,16-bit word &
fixed-size partitions of 128
\\

{\opf} &
cardinality &
no &
\,\,\,32-bit word &
fixed-size partitions of 128
\\

{\simples} &
cardinality &
no &
\,\,\,64-bit word &
fixed-size partitions of 128
\\

{\qmx} &
cardinality &
yes &
128-bit word &
fixed-size partitions of 128
\\

{\roaropt} &
universe &
yes &
byte &
single-span
\\

{\slice} &
universe &
yes &
byte &
multi-span
\\

\bottomrule
\end{tabular}}
\label{tab:strategies}
\end{table}

\begin{table}[t]
\caption{The datasets used in the experiments.}
\subfloat[basic statistics]{
\scalebox{0.8}{\begin{tabular}{lrrr}
\toprule

& {\gov} & {\clue}  & {\cc} \\

\midrule


\small{Lists}
  & \num{39177}
  & \num{96722}
  & \num{76474}
  \\
  
\small{Universe}
  & \num{24622347}
  & \num{50131015}
  & \num{43530315}
  \\

\small{Integers}
  & \num{5322883266}
  & \num{14858833259}
  & \num{19691599096}  
  \\

\small{Entropy of the gaps}
	& 3.02
  	& 4.46
  	& 5.44
  	\\

\small{$\lceil \log_2 \rceil$ of the gaps}
	& 1.35
  	& 2.28
  	& 2.99
  	\\
  	  	
\bottomrule
\end{tabular}}
\label{tab:datasets_a}
}
\subfloat[TREC 2005/06 queries]{
\scalebox{0.8}{
\begin{tabular}{lccc}
\toprule

& \multicolumn{1}{c}{\gov}
& \multicolumn{1}{c}{\clue} 
& \multicolumn{1}{c}{\cc} \\

\midrule

\small{Queries}
  & {34,327} 
  & {42,613} 
  & {22,769} 
  \\
  
\small{2 \,\, terms}
  & 32.2\%
  & 33.6\%
  & 37.5\%
  \\
  
\small{3 \,\, terms}
  & 26.8\%
  & 26.5\%
  & 27.3\%
  \\

\small{4 \,\, terms}
  & 18.2\%
  & 17.7\%
  & 16.8\%
  \\

\small{5+ terms}
  & 22.8\%
  & 22.2\%
  & 18.4\%
  \\


\bottomrule
\end{tabular}}
}
\vspace{-0.5cm}
\label{tab:datasets}
\end{table}

\parag{Datasets} We perform the experiments on the following
standard test collections.
{\gov} is the TREC 2004 Terabyte Track test collection,
consisting in roughly 25 million \textsf{.gov} sites crawled in early 2004.
The documents are truncated to 256 KB.
{\clue} is the ClueWeb 2009 TREC Category B test collection,
consisting in roughly 50 million English web pages
crawled between January and February 2009.
{\cc} is a dataset of news freely available from
\href{http://commoncrawl.org/2016/10/news-dataset-available}{CommonCrawl}.
Precisely, the dataset consists of the news appeared from 09/01/16 to 30/03/18.

Identifiers were assigned to documents according to the lexicographic order
of their URLs~\cite{2007:silvestri} (see also the discussion
at the beginning of Section~\ref{sec:list}).
From the original collections we retain all lists whose size
is larger than 4096. The postings belonging to these lists
cover 93\%, 94\%, and 98\% of the total postings of
{\gov}, {\clue}, and {\cc} respectively.
From the TREC 2005 and TREC 2006 Efficiency
Track topics, we selected all queries whose terms
are in the lexicons of the tested collection.
Table~\ref{tab:datasets} reports the statistics for the collections.

\parag{Experimental setting and methodology}
Experiments are performed on a server machine equipped with
Intel i9-9900K cores (@3.6 GHz), 64 GB of RAM DDR3 (@2.66 GHz)
and running Linux 5 (64 bits).
Each core has two private levels of cache memory: 32 KiB L1 cache
(one for instructions and one for data);
256 KiB for L2 cache. A shared L3 cache spans 16,384 KiB.


The whole code is written in C++ and
compiled with \texttt{gcc} 9.2.1 using the highest
optimization setting, i.e., with compilation flags \texttt{-O3}
and \texttt{-march=native}.


We build the indexes in internal memory and write the corresponding
data structures to a file on disk.
To perform the queries, the data structure is memory mapped from the file
and a warming-up run is executed to fetch the necessary pages from disk.
To test the speed of intersection and union,
we use a random sampling of 1000 queries for each number of query terms
from 2 to 5+ (with 5+ meaning queries with \emph{at least} 5 terms).
Each experiment was repeated 3 times
to smooth fluctuations during measurements.
The time reported is the average among these runs.


\begin{table}
\caption{Space effectiveness in total {\total} and bits per integer,
and nanoseconds per decoded integer.}
\scalebox{\mytablescale}{\begin{tabular}{
l
SSS
l
SSS
l
SSS
}
\toprule

\multirow{2}{*}{Method}
& \multicolumn{3}{c}{\gov}
&
& \multicolumn{3}{c}{\clue}
&
& \multicolumn{3}{c}{\cc}
\\

\cmidrule(lr){2-4}
\cmidrule(lr){6-8}
\cmidrule(lr){10-12}

& {\total} & {\bpi} & {\nspi}
&
& {\total} & {\bpi} & {\nspi}
&
& {\total} & {\bpi} & {\nspi}
\\

\midrule

{\vb}
& 5.4601 & 8.81151 & 0.958686 
&
& 15.9200 & 9.20342 & 1.0906 
&
& 21.2899 & 9.28716 & 1.03278
\\

{\optvb}
& 2.4087 & 3.88721 & 0.734872 
&
& 9.8923 & 5.71879 & 0.92005 
&
& 14.7271 & 6.42431 & 0.720593
\\

{\interp}
& 1.8230 & 2.94206 & 5.06396
&
& 7.6626 & 4.42979 & 6.31163
&
& 12.0198 & 5.24335 & 6.96839
\\

{\cdelta} 
& 2.3170 & 3.73914 & 3.56276 
&
& 8.9493 & 5.17367 & 3.71787 
&
& 14.5840 & 6.36192 & 3.84854
\\



{\rice}
& 2.53072 & 4.08402 & 2.92386 
&
& 9.17985 & 5.3069 & 3.24619 
&
& 13.3445 & 5.82117 & 3.32222
\\

{\opt}
& 1.9310 & 3.11627 & 0.764215
&
& 8.6263 & 4.98693 & 1.09899
&
& 12.4955 & 5.45084 & 1.30799
\\

{\dint}
& 2.1854 & 3.52686 & 1.12957 
&
& 9.2578 & 5.35198 & 1.5596 
&
& 14.7585 & 6.438 & 1.65251
\\

{\opf}
& 2.2468 & 3.62592 & 1.37746 
&
& 9.4526 & 5.46462 & 1.78855 
&
& 13.9180 & 6.07138 & 1.52592
\\

{\simples}
& 2.5943 & 4.18663 & 1.52539 
&
& 10.1258 & 5.85379 & 1.87015 
&
& 14.6837 & 6.40541 & 1.8917
\\

{\qmx}
& 3.1731 & 5.12074 & 0.795423 
&
& 12.6017 & 7.28511 & 0.86883 
&
& 16.9570 & 7.39705 & 0.83607
\\

{\roaropt}
& 4.1070 & 6.627809 & 0.496726
&
& 16.9230 & 9.783261 & 0.707462
&
& 21.7474 & 9.486765 & 0.608148
\\

{\slice}
& 2.6707 & 4.31 & 0.527622
&
& 12.2094 & 7.05834 & 0.684622
&
& 17.8313 & 7.77846 & 0.691265
\\

\bottomrule
\end{tabular}}
\label{tab:space_and_decoding}
\end{table}

\parag{Compression effectiveness}
In Table~\ref{tab:space_and_decoding} we report the 
compression effectiveness of each method expressed as
total {\total} and bit-per-integer rate.
The following considerations hold
pretty much consistently across the three tested datasets.
The most effective method is {\bic} with {\pef} being
close second. Observe that both methods come very close to
the entropy of gaps (with {\bic} being even better), as reported
if Table~\ref{tab:datasets_a}.
The least effective methods are {\vbyte} and {\roaropt}
(in particular, {\roaropt} is sensibly better than {\vbyte} on {\gov}
but performs worse on the other two datasets).
The representations
{\optvb}, {\cdelta}, {\rice}, {\dint}, {\opf} and {\simples}
are all similar in space,
taking roughly 3 -- 4, 5 -- 6
and 6 -- 6.5 {\bpi}
for {\gov}, {\clue}, and {\cc} respectively.
The {\qmx} and {\slice} approaches stand in a middle position
between the former two classes of methods.


\parag{Sequential decoding}
Table~\ref{tab:space_and_decoding} also reports
the average nanoseconds
spent per decoded integer, measured after decoding
all lists in the index.
For all the different methods, the result of decoding
a list
is materialized into an output buffer of 32-bit integers.
Again, results are consistent across the different datasets.

The fastest methods are {\roaropt} and {\slice}
thanks to their ``simpler'' design
involving byte-aligned codes, bitmaps,
and the use of SIMD instructions, allowing a value
to be decoded in 0.5 -- 0.7 nanoseconds.
The methods {\optvb}, {\qmx}, and {\pef}
are the second fastest, requiring 0.7 -- 1.3 nanoseconds on average.
In particular, {\optvb} and {\pef} gain most of their
speed thanks to the efficient decoding of dense bitmaps.
The methods {\bic}, {\cdelta}, and {\rice} are the slowest as
they only decode one symbol at a time
(observe that {\bic} is almost $2\times$ slower than the other two
because of its recursive implementation).
The other mechanisms {\vbyte}, {\dint}, {\opf} and {\simples}
provide similar efficiency, on average decoding an integer in
1 -- 1.9 nanoseconds.

Lastly, recall that all methods -- except {\bic}, {\pef}, {\roaropt},
and {\slice} -- require a prefix-sum computation
because they encode the gaps between the integers.
In our experiments, we determined that the
cost of computing the prefix-sum of the gaps is 0.5 nanoseconds per integer.
This cost, sometimes, dominates that of decoding the gaps.

\begin{figure}[t]
\includegraphics[scale=0.75]{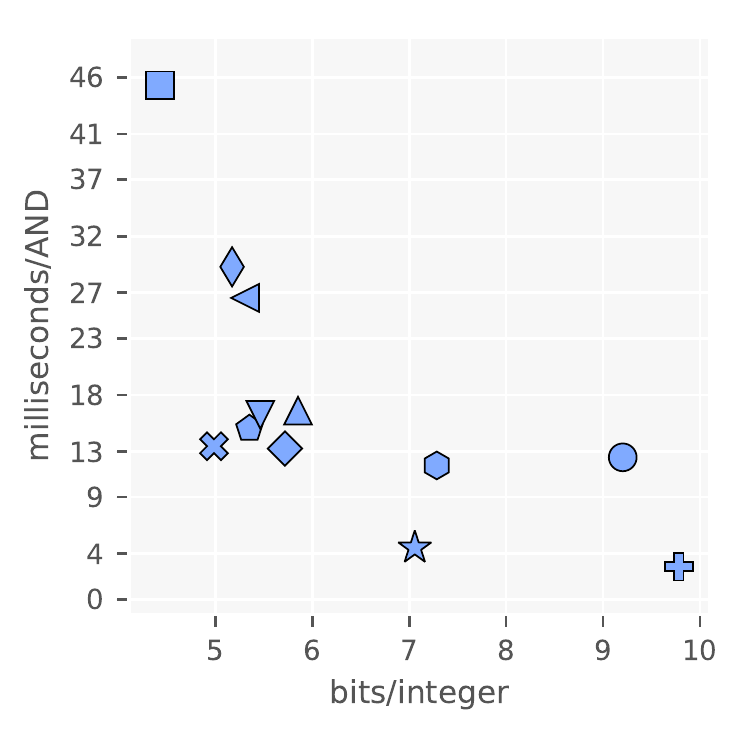}
\includegraphics[scale=0.75]{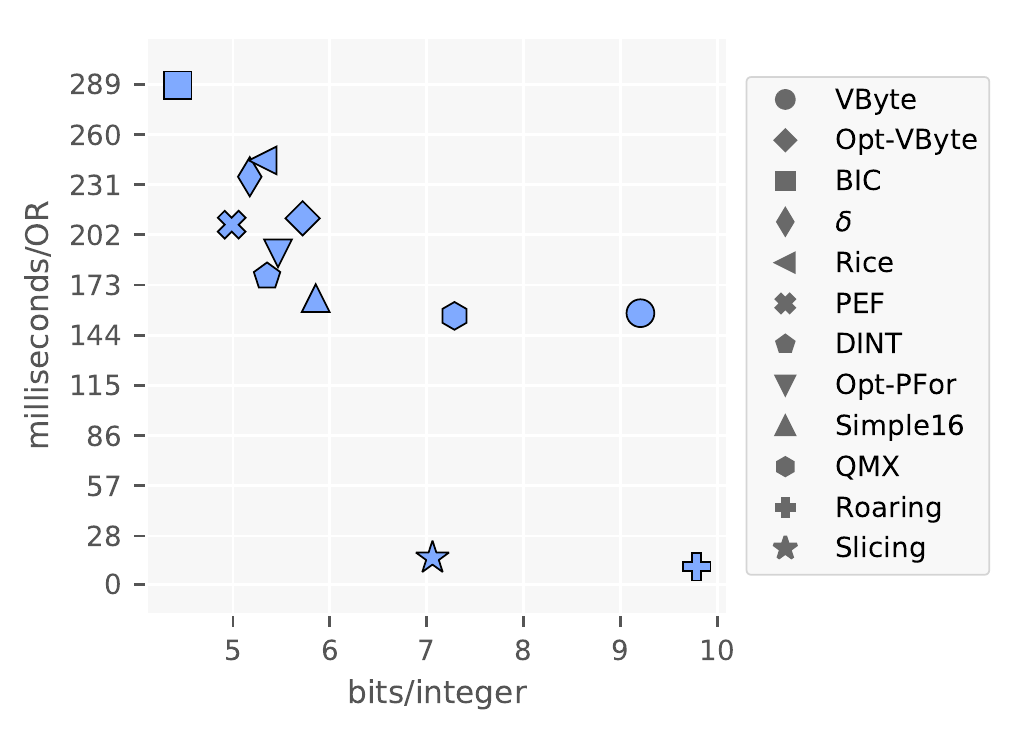}
\caption{Space/time trade-off curves for the {\clue} dataset.
\label{fig:pareto}}
\end{figure}

\parag{Boolean AND/OR queries}
We now consider the operations of list intersection and union.
Table~\ref{tab:AND} and~\ref{tab:OR} report
the timings by varying the number of query terms.
Fig.~\ref{fig:pareto} displays the data in the tables
for the {\clue} dataset
along space/time trade-off curves,
(thus, also incorporating
the space information brought by Table~\ref{tab:space_and_decoding})
and with the time being the ``avg.'' column.
Almost identical shapes were
obtained for the other datasets.
When considering the general trade-off, especially
highlighted by the plots, we see that
the trend of the trade-off
is the same for both intersections
and unions, even across three different datasets.
Therefore we can make some general points.


For methods partitioned by cardinality,
the efficiency of intersection is
strictly correlated to that of
$\Nextgeq(x)$, an operation returning the
smallest integer $z \geq x$;
the efficiency of
union is correlated to that of
sequential decoding.
This is not necessarily true for
{\roaropt} and {\slice} that, being partitioned
by universe rather than cardinality, employ
an intersection algorithm that does not use
{\Nextgeq}, nor a merging algorithm that
loop through every single integer in a sequence.

There is a cluster of techniques providing
similar efficiency/effectiveness trade-offs,
including {\pef}, {\dint}, {\optvb}, {\simples},
{\opf} and {\qmx}, whereas
{\bic}, {\cdelta} and {\rice} are always the slowest
and dominated by the aforementioned techniques.

The procedures employed by
{\roaropt} and {\slice} outperform in efficiency
all techniques by a wide margin.
Again, this is possible because they do not
consider one-symbol-at-a-time operations,
rather they rely on the intrinsic parallelism
of inexpensive bitwise instructions over 64-bit words. 
The difference in query time between {\roaropt} and
{\slice} has to be mostly attributed to the
SIMD instructions that are better exploited by {\roaropt}
thanks to its simpler design.
To confirm this, we performed an experiment
with SIMD disabled and obtained almost identical timings
to those of {\slice}.
However, these representations
take more space than the aforementioned
cluster of techniques: {\slice} stands
in a middle position between such
cluster and {\roaropt}.

Also observe that, for {\AND} queries,
the efficiency gap between the methods
partitioned by universe and the ones partitioned by cardinality
reduces when more query terms are considered.
This is because the queries becomes progressively
more selective (on average), hence allowing large skips
to be performed by {\Nextgeq}.
On the contrary, methods partitioned by universe
only skip at a coarser level, e.g., chunks
containing at most $2^{16}$ integers, therefore
the cost for in-chunk calculations is always paid,
even when only few integers belong to the result set.
Note that this is not true for {\OR} queries:
the gap becomes progressively more evident
with more query terms.



\begin{table}[t]
\caption{Milliseconds spent per {\AND} query by varying the number of query terms.
\label{tab:AND}}
\scalebox{0.8}{\sisetup{
round-precision = 1
}
\renewcommand{\tabcolsep}{0.2mm}
\begin{tabular}{
l@{ }
S[table-number-alignment=center]
S[table-number-alignment=center]
S[table-number-alignment=center]
S[table-number-alignment=center]
S[table-number-alignment=center]
l@{ }
S[table-number-alignment=center]
S[table-number-alignment=center]
S[table-number-alignment=center]
S[table-number-alignment=center]
S[table-number-alignment=center]
l@{ }
S[table-number-alignment=center]
S[table-number-alignment=center]
S[table-number-alignment=center]
S[table-number-alignment=center]
S[table-number-alignment=center]
}
\toprule

\multirow{2}{*}{Method}

& \multicolumn{5}{c}{\gov}
&
& \multicolumn{5}{c}{\clue}
&
& \multicolumn{5}{c}{\cc}
\\

\cmidrule(lr){2-6}
\cmidrule(lr){8-12}
\cmidrule(lr){14-18}

& 2 & 3 & 4 & \multicolumn{1}{c}{5+} & \multicolumn{1}{c}{avg.}
&
& 2 & 3 & 4 & \multicolumn{1}{c}{5+} & \multicolumn{1}{c}{avg.}
&
& 2 & 3 & 4 & \multicolumn{1}{c}{5+} & \multicolumn{1}{c}{avg.}
\\

\midrule

{\vb}
& 2.187190 & 2.805776 & 2.729091 & 3.302966 & 2.75625575
&
& 10.236636 & 12.141142 & 13.748690 & 13.876838 & 12.5008265 
&
& 14.030115 & 22.440629 & 19.717454 & 21.902058 & 19.522564 
\\

{\optvb}
& 2.757718 & 3.138205 & 2.815726 & 3.194730 & 2.97659475 
&
& 12.239734 & 13.274988 & 13.974772 & 13.607428 & 13.2742305 
&
& 15.985631 & 23.181998 & 19.605050 & 20.274870 & 19.76188725 
\\

{\interp}
& 6.789404 & 9.730492 & 10.376179 & 13.159253 & 10.013832 
&
& 31.683833 & 44.219982 & 51.465445 & 53.821621 & 45.29772025 
&
& 45.574952 & 79.737292 & 76.858444 & 88.838230 & 72.7522295 
\\

{\cdelta} 
& 4.554581 & 6.279071 & 6.514751 & 8.234386 & 6.39569725 
&
& 20.903358 & 28.329439 & 33.475450 & 34.462521 & 29.292692 
&
& 28.606313 & 50.930272 & 47.963941 & 55.577109 & 45.76940875 
\\

{\rice}
& 4.103047 & 5.619431 & 5.813134 & 7.307773 & 5.71084625 
&
& 19.220993 & 25.705405 & 30.223074 & 31.053789 & 26.55081525 
&
& 26.503436 & 46.474844 & 43.499346 & 50.093738 & 41.642841 
\\

{\opt}
& 2.538159 & 3.063740 & 2.833232 & 3.220635 & 2.9139415 
&
& 12.293293 & 13.466791 & 14.357337 & 13.803540 & 13.48024025 
&
& 17.231502 & 24.607306 & 20.962324 & 21.887795 & 21.17223175 
\\

{\dint}
& 2.482701 & 3.302166 & 3.269413 & 4.068720 & 3.28075 
&
& 11.897265 & 14.626435 & 16.486294 & 17.136501 & 15.03662375 
&
& 16.912144 & 27.281171 & 24.583464 & 28.052237 & 24.207254 
\\

{\opf}
& 2.607119 & 3.457250 & 3.471663 & 4.275161 & 3.45279825 
&
& 12.757529 & 15.935078 & 17.992163 & 18.318734 & 16.250876 
&
& 16.561864 & 27.241049 & 24.300965 & 27.143666 & 23.811886 
\\

{\simples}
& 2.787031 & 3.694952 & 3.701827 & 4.569633 & 3.68836075
&
& 12.849040 & 16.294048 & 18.438184 & 18.866284 & 16.611889 
&
& 17.614961 & 28.758527 & 26.258377 & 29.472442 & 25.52607675
\\

{\qmx}
& 2.002297 & 2.553365 & 2.501374 & 3.044888 & 2.525481 
&
& 9.615046 & 11.487097 & 12.955751 & 13.078712 & 11.7841515 
&
& 13.322526 & 21.491361 & 18.784743 & 20.811573 & 18.60255075
\\

{\roaropt}
& 0.314378 & 0.536833 & 0.680153 & 0.835483 & 0.59171175
&
& 1.549764 & 2.522875 & 3.119923 & 4.303670 & 2.874058
&
& 1.140922 & 2.028363 & 2.578873 & 4.056526 & 2.451171
\\

{\slice}
& 0.291648 & 1.046495 & 1.206713 & 1.584944 & 1.03245 
&
& 1.477132 & 4.492650 & 5.391111 & 6.743450 & 4.52608575 
&
& 1.813670 & 4.328495 & 5.050425 & 6.018512 & 4.3027755
\\

\bottomrule
\end{tabular}}
\end{table}

\begin{table}[t]
\caption{Milliseconds spent per {\OR} query by varying the number of query terms.
\label{tab:OR}}
\scalebox{0.8}{\sisetup{
round-precision = 1
}
\renewcommand{\tabcolsep}{0.2mm}
\begin{tabular}{
l@{ }
S[table-number-alignment=center]
S[table-number-alignment=center]
S[table-number-alignment=center]
S[table-number-alignment=center]
S[table-number-alignment=center]
l@{ }
S[table-number-alignment=center]
S[table-number-alignment=center]
S[table-number-alignment=center]
S[table-number-alignment=center]
S[table-number-alignment=center]
l@{ }
S[table-number-alignment=center]
S[table-number-alignment=center]
S[table-number-alignment=center]
S[table-number-alignment=center]
S[table-number-alignment=center]
}
\toprule

\multirow{2}{*}{Method}

& \multicolumn{5}{c}{\gov}
&
& \multicolumn{5}{c}{\clue}
&
& \multicolumn{5}{c}{\cc}
\\

\cmidrule(lr){2-6}
\cmidrule(lr){8-12}
\cmidrule(lr){14-18}

& 2 & 3 & 4 & \multicolumn{1}{c}{5+} & \multicolumn{1}{c}{avg.}
&
& 2 & 3 & 4 & \multicolumn{1}{c}{5+} & \multicolumn{1}{c}{avg.}
&
& 2 & 3 & 4 & \multicolumn{1}{c}{5+} & \multicolumn{1}{c}{avg.}
\\

\midrule

{\vb}
& 6.805895 & 24.413179 & 54.749224 & 131.659482 & 54.406945
&
& 20.064903 & 71.331185 & 156.035270 & 379.527670 & 156.739757 
&
& 24.374824 & 94.498804 & 178.816094 & 391.388465 & 172.26954675 
\\

{\optvb}
& 11.003441 & 35.672215 & 77.377703 & 176.012789 & 75.016537
&
& 31.315430 & 101.387425 & 213.382186 & 500.139796 & 211.55620925 
&
& 36.356127 & 128.025115 & 231.984099 & 510.368780 & 226.68353025 
\\

{\interp}
& 16.650048 & 50.304635 & 104.994259 & 238.794393 & 102.68583375
&
& 49.885611 & 145.264443 & 290.448272 & 668.157821 & 288.43903675 
&
& 64.440340 & 193.764301 & 332.587262 & 692.451685 & 320.810897
\\

{\cdelta} 
& 12.568083 & 40.784598 & 87.929234 & 202.498092 & 85.94500175 
&
& 34.914084 & 112.949214 & 236.700915 & 557.705287 & 235.567375 
&
& 42.161067 & 144.925453 & 263.792565 & 571.251974 & 255.53276475
\\

{\rice}
& 13.442832 & 43.090598 & 93.315179 & 211.323788 & 90.29309925
&
& 36.814944 & 118.223476 & 248.531221 & 576.551637 & 245.0303195 
&
& 43.635286 & 149.269469 & 270.530053 & 585.551783 & 262.24664775
\\

{\opt}
& 10.173077 & 32.977762 & 71.700607 & 164.247834 & 69.77482 
&
& 31.091084 & 99.672703 & 208.515157 & 492.308774 & 207.8969295 
&
& 37.565794 & 127.548571 & 232.605817 & 507.123582 & 226.210941 
\\

{\dint}
& 8.546517 & 28.525952 & 63.747554 & 147.632307 & 62.1130825
&
& 24.880267 & 84.096185 & 178.844496 & 424.291327 & 178.02806875 
&
& 30.560057 & 109.188208 & 200.372893 & 432.678240 & 193.1998495
\\

{\opf}
& 8.949865 & 31.062088 & 69.411155 & 161.383393  & 67.70162525
&
& 27.031952 & 90.770767 & 193.964631 & 453.480761 & 191.31202775 
&
& 31.303298 & 113.150031 & 208.980109 & 447.204679 & 200.15952925
\\

{\simples}
& 7.752507 & 26.225273 & 58.254741 & 138.238449 & 57.6177425
&
& 23.668365 & 78.005275 & 165.530239 & 394.667560 & 165.46785975 
&
& 28.744876 & 101.539389 & 185.340008 & 397.776452 & 178.35018125 
\\

{\qmx}
& 6.641653 & 23.842591 & 53.403706 & 128.110673 & 52.99965575
&
& 19.692581 & 70.030373 & 153.153887 & 377.873294 & 155.18753375 
&
& 23.971722 & 92.619244 & 175.239619 & 382.378001 & 168.5521465 
\\

{\roaropt}
& 1.182470 & 2.753535 & 4.310408 & 6.406141 & 3.6631385
&
& 4.673867 & 8.965774 & 12.023105 & 15.711115 & 10.34346525
&
& 3.834272 & 7.607278 & 10.451422 & 15.056181 & 9.23728825
\\

{\slice}
& 1.297511 & 4.033000 & 6.258782 & 9.197781 & 5.1967685 
&
& 4.970250 & 12.828265 & 18.129565 & 25.266133 & 15.29855325 
&
& 5.840060 & 12.906089 & 17.312807 & 22.994044 & 14.76325
\\

\bottomrule
\end{tabular}}
\end{table}

\section{Conclusions and future research directions}\label{sec:conclusions}

The problem of introducing a compression format
for sorted integer sequences, with good practical 
intersection/union performance,
is well-studied and important, given
its fundamental application to large-scale retrieval
systems such as Web search engines.
For that reason, inverted index compression is still a very 
active field of research that began several decades ago.
With this article, we aimed at surveying the encoding
algorithms suitable to solve the problem.
However, electing a solution as the ``best'' one
is not generally easy, rather the many space/time trade-offs
available can satisfy different application requirements
and the solution should always be determined
by considering the actual data distribution.
To this end, we also offer an experimental
comparison between
many of the techniques described in this article.
The different space/time trade-offs
assessed by this analysis are summarized by Fig.~\ref{fig:pareto}.

Because of the maturity reached by the state-of-the-art
and the specificity of the problem,
identifying future research directions is not immediate.
We mention some promising ones.
In general, devising ``simpler'' compression
formats that can be decoded with algorithms using
low-latency instructions (e.g., bitwise) and with as few
branches as possible, is a profitable line of research,
as demonstrated by the experimentation in this article.
Such algorithms favour the super-scalar execution of modern CPUs
and are also suitable for SIMD instructions.
Another direction could look at devising
\emph{dynamic and compressed} representations for integer
sequences, able of also supporting additions and deletions.
This problem is actually a specific case of the more general
\emph{dictionary problem}, which is a fundamental
textbook problem.
While a theoretical solution already exists
with all operations supported in optimal time
and compressed space~\cite{PV17CPM},
an implementation with good practical performance could
be of great interest for dynamic inverted indexes.

\begin{acks}
The authors are grateful to Daniel Lemire, Alistair Moffat,
Giuseppe Ottaviano, Matthias Petri, Sebastiano Vigna,
and the anonymous referees
for having carefully read earlier versions of the manuscript.
Their valuable suggestions substantially improved the quality of exposition,
shape, and content of the article.

This work was partially supported by the BIGDATAGRAPES (EU H2020 RIA, grant agreement N\textsuperscript{\b{o}}780751), the ``Algorithms, Data Structures and Combinatorics for Machine Learning'' (MIUR-PRIN 2017), and the OK-INSAID (MIUR-PON 2018, grant agreement N\textsuperscript{\b{o}}ARS01\_00917) projects.
\end{acks}

\bibliographystyle{ACM-Reference-Format}
\bibliography{bibliography}

\end{document}